\documentclass[aps,prl,showpacs,superscriptaddress,amsmath,amssymb,footinbib,
reprint,
floatfix,
]{revtex4-1}
\usepackage{graphicx}
\usepackage[hypertexnames=false]{hyperref}
\begin{document}
%
%
\title{Temperature-Dependent Lifetimes of Low-Frequency Adsorbate Modes from Non-Equilibrium Molecular Dynamics Simulations }
\author{Francesco Nattino}
\email{f.nattino@chem.leidenuniv.nl}
\altaffiliation[\textit{Present address:} ]{Theory and Simulations of Materials (THEOS) and National Centre for Computational Design and Discovery of Novel Materials (MARVEL), \'{E}cole Polytechnique F\'{e}d\'{e}rale de Lausanne, CH-1015 Lausanne, Switzerland}
\affiliation{Leiden Institute of Chemistry, Leiden University, Gorlaeus Laboratories, P.O. Box 9502, 2300 RA Leiden, The Netherlands}
\author{J\"{o}rg Meyer}
\email[Corresponding author: ]{j.meyer@chem.leidenuniv.nl}
\affiliation{Leiden Institute of Chemistry, Leiden University, Gorlaeus Laboratories, P.O. Box 9502, 2300 RA Leiden, The Netherlands}
\date{\today} 
\begin{abstract}
We present calculations on the damping of a low-frequency adsorbate mode on a metal surface, namely the frustrated translation of Na on Cu(100). For the first time, vibrational lifetimes of excited adlayers are extracted from non-equilibrium molecular dynamics calculations accounting for both the phononic and the electronic dissipation channels. The relative contributions of the two damping mechanisms, which we show to be additive, are found to disagree with textbook predictions. A simple model based on separable harmonic and anharmonic contributions is able to semi-quantitatively reproduce the temperature dependence of the computed lifetimes.
\end{abstract}
\maketitle
%

Metal-based catalysts are widely employed in the chemical and energetic sectors \cite{Somorjai-PNAS-2011}. Their role consists in providing a substrate for adsorption, diffusion and transformation of the reactants into the products, which eventually desorb from the catalyst surface \cite{Nilsson-ChemicalBondingAtSurfacesAndInterfaces-2008}. The efficiency of the catalyst is crucially affected by the energy exchange between the reaction intermediates and the substrate, which on metal surfaces is ruled by the competition between phonon emission and electron-hole pair excitation. 
Disentangling the contribution of these two damping mechanisms is therefore extremely relevant for the fundamental understanding of dissipation processes at the atomic scale. 

Small adsorbates on single-crystal metal surfaces represent ideal systems to investigate the  energy exchange with the underlying substrate, as their vibrational damping can be accurately characterized e.g. through helium scattering experiments \cite{Graham-PRL-1997, Graham-SurfSciRep-2003}. For adsorbate modes whose frequency $\omega$ falls well within the substrate phonon spectrum ($\omega<10$ meV), textbook knowledge predicts vibrational lifetimes to be dominated by phonon excitations \cite{Graham-SurfSciRep-2003, Politano-SurfSciRep-2013}. According to this picture, the electronic contribution to the lifetime of the frustrated translation (FT) mode for alkali metal adatoms on transition metal surfaces is expected to be negligible \cite{Graham-SurfSciRep-2003, Politano-SurfSciRep-2013}. This expectation stands in clear contrast with the results of a recent study that focused on sodium atoms diffusing on Cu(111) \cite{Rittmeyer-PRL-2016}. In fact, calculations that followed theoretical developments \cite{Juaristi-PRL-2008, BlancoRey-PRL-2014, Rittmeyer-PRL-2015} of the dynamics beyond the Born-Oppenheimer approximation \cite{Tully-JVacSciTechnol-1993, Tully-JCP-1980, HeadGordon-JCP-1995} have estimated the electronic dissipation mechanism to be as relevant as $\approx 25 \%$ of the  experimentally-determined friction coefficient \cite{Rittmeyer-PRL-2016}. This is despite the dispersionless FT-mode of Na adatoms on low-index Cu surfaces should very efficiently couple to bulk modes due to the large frequency overlap with the long-wavelength region of the substrate phonon continuum ($\hbar\omega \approx 6$~meV for Na on Cu(100) \cite{Graham-PRL-1997, Politano-SurfSciRep-2013}). 


Unfortunately, the phononic counterpart to the electronic channel in the vibrational damping of adsorbate modes is not straightforward to evaluate in a quantitative way \cite{Chen-Science-2018}. On the one hand, perturbative approaches relies on the evaluation of the system's third-order force constants \cite{Maradudin-PhysRev-1962}, which typically involves a significant computational effort. On the other hand, molecular dynamics (MD) techniques require large simulation cells in order to represent bulk-projected phonons with a sufficient resolution over the surface Brillouin zone (SBZ) \cite{Meyer-PhDThesis, Meyer-Angew-2014}. In this scenario, the state-of-the-art method for estimating phonon-dominated vibrational lifetimes remains the elastic continuum model (ECM) initially proposed by Persson and Ryberg \cite{Persson-PRB-1985} and further developed for periodic adlayers by Rappe \textit{et al.} \cite{Lewis-JCP-1998, Pykhtin-PRL-1998}. This model is based on a continuum description of the substrate and it allows to estimate adsorbate mode lifetimes from a very limited set of system-specific parameters. Despite the simplicity of the ECM, Rappe and coworkers showed that its periodic-adlayer generalization could semi-quantitatively reproduce measured coverage-dependent lifetimes for the carbon monoxide FT-mode on Cu(100) \cite{Lewis-JCP-1998, Pykhtin-PRL-1998}.

In this letter, we present an approach that has allowed us for the first time to directly extract adsorbate-mode lifetimes from non-equilibrium simulations. The low computational cost of classical MD calculations enables the atomistic description of large simulation cells, so that a dense sampling of the substrate SBZ is naturally obtained. At the same time, coverage and temperature effects can be realistically modeled. We have focused here on the damping of the FT-mode of Na adatoms on Cu(100), exploiting this prototypical system, relevant to catalysis \cite{Lackey-JCSFT1-1987}, to test the validity of the approximations underlying the ECM. We have found a pronounced temperature dependence for the Na FT-mode lifetime, in agreement with experimental trends \cite{Graham-PRL-1997, Politano-SurfSciRep-2013}. Moreover, by accounting for electron-phonon coupling via Langevin electronic friction dynamics \cite{Tully-JCP-1980, Tully-JVacSciTechnol-1993, HeadGordon-JCP-1995, Juaristi-PRL-2008} we have found a large electronic contribution to the lifetime, in agreement with the recent simulations from Rittmeyer \textit{et al.} \cite{Rittmeyer-PRL-2016}. Most importantly, our study produces valuable insight into the harmonic and anharmonic contributions to the phonon-mediated adsorbate damping and to the additivity of the electronic and the phononic dissipation channels.

We base our calculations on a specifically-constructed potential of the embedded atom method (EAM) form \cite{Daw-PRL-1983}, which very accurately reproduces the experimentally measured surface phonon band structure for Na on Cu(100) \cite{Senet-ChemPhysLett-1999, Graham-SurfSciRep-2003, SM}. Furthermore, it also yields excellent agreement with the Na-Cu interaction potential obtained from helium atom scattering experiments \cite{Graham-PRL-1997, Graham-PRB-1997-1, SM}. Based on this potential we perform classical molecular dynamics (MD) simulations with the LAMMPS code \cite{Plimpton-JCompPhys-1995} for different Na-adlayer structures (\textit{vide infra}) and including temperature by pre-equilibrating the entire system \cite{West-PRB-2007,SM}. In order to ensure a proper theory-theory comparison, we use the formulation of the ECM for periodic adlayers by Rappe \textit{et al.} \cite{Lewis-JCP-1998, Pykhtin-PRL-1998} and obtain all parameters required (bulk lattice constant, density, transversal as well as longitudinal speeds of sound and adsorbate frequency) directly from bulk and surface calculations with our EAM potential \cite{SM}. 

Consistent with the previous work on the ECM and representative for experimental conditions encountered in optical pump-probe experiments \cite{watanabe2005,fuyuki2007}, we initially excite a FT 
by depositing an amount of energy larger than the thermal background into the translational degrees of freedom of the Na adatoms
, such that all of them initially vibrate in phase along the same 
crystallographic direction ($\bar{\Gamma}$ excitation). Constant energy MD calculations for these non-equilibrium initial conditions (NEMD) combined with a phonon mode projection scheme \cite{McGaughey-PRB-2004, Meyer-PhDThesis} allow us to follow the equilibration of the small amount of excess energy within the phonon bath. Inspired by the work of Estreicher \textit{et al.} on the vibrational lifetimes of bulk defects \cite{West-PRL-2006}, we can accurately quantify the lifetime $\tau_{\mathrm{NEMD}}$ by fitting an exponential decay function to the energy in the projected phonon mode corresponding to the FT after averaging over 100 trajectories \cite{SM}. 
Compared to the pioneering work of Tully \cite{Tully-JCP-1980, Tully-JVacSciTechnol-1993}, which has relied on time correlation functions of a harmonic system at equilibrium, our approach is much more direct and more accurate.

\begin{figure}
\includegraphics[width=8.6cm]{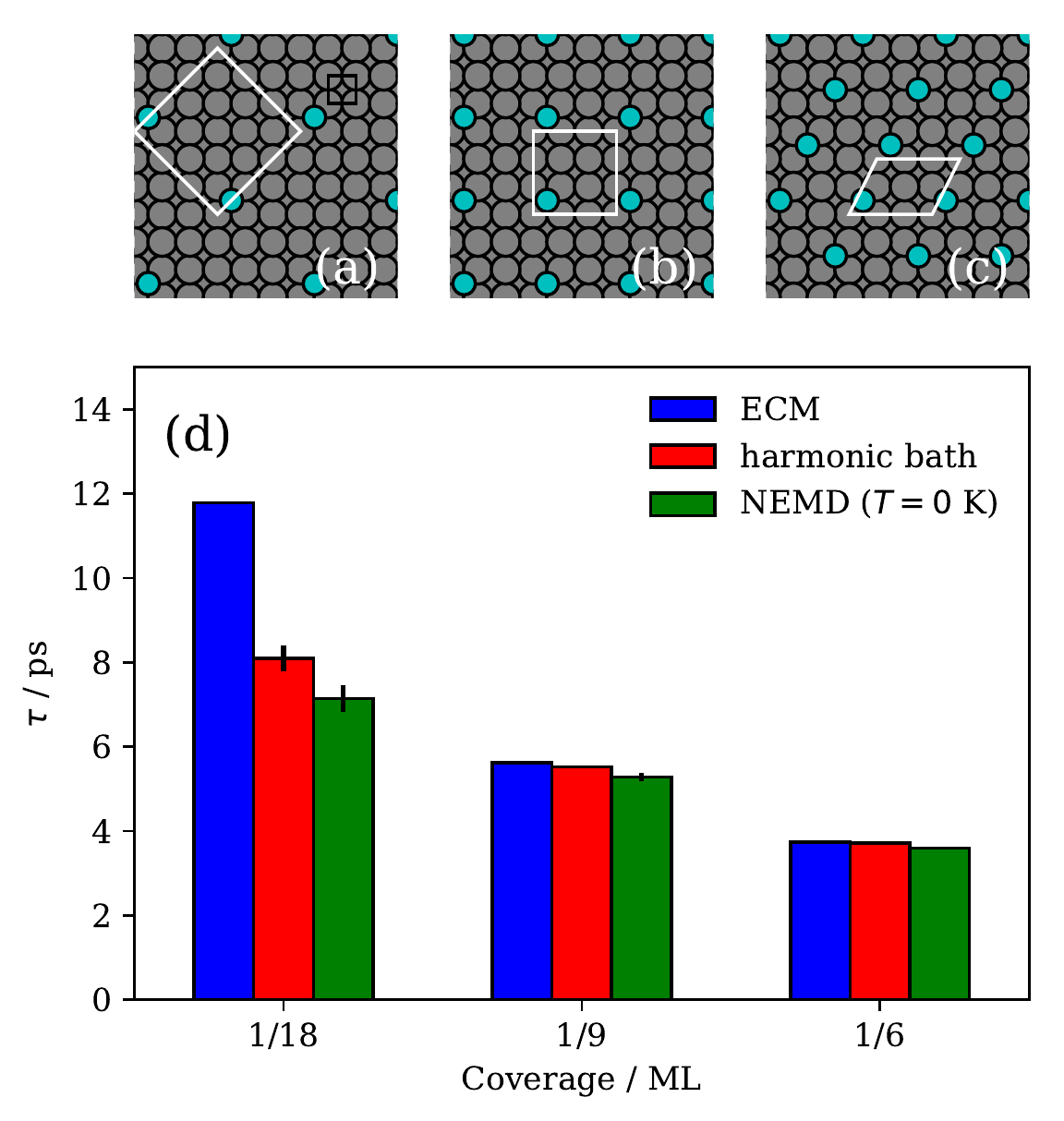}
\caption{\label{fig:effect_of_coverage}%
(color online). (a-c) Illustrations representing the three overlayer structures considered in this work.
(a) c$(6\times6)$, (b) p$(3\times3)$ and(c) $\left(\begin{array}{cc} 3 & 0 \\ 1 & 2 \end{array}\right)$ Na-adlayer structures corresponding to coverages $\theta=1/18$, $\theta=1/9$ and $\theta=1/6$ given in fractions of a monolayer (ML), respectively. Copper and sodium atoms are shown in gray and cyan, respectively. The adsorbate unit cells are shown in white and compared to the Cu(100) primitive cell, shown in black in (a). (d) Na FT-mode lifetimes calculated for the three overlayer structures. Blue, red and green bars correspond to the values obtained with the ECM, the harmonic bath and our NEMD model (at $T=0$ K), respectively. Error bars reflect the quality of the fits employed to estimate the lifetimes \cite{SM}.}
\end{figure}

We have simulated three overlayer structures, illustrated in Fig. \ref{fig:effect_of_coverage} (a)-(c): the experimentally observed $\left(\begin{array}{cc} 3 & 0 \\ 1 & 2 \end{array}\right)$ adsorption structure \cite{Graham-PRB-1997-2} that corresponds to a surface coverage $\theta=1/6\approx16.7\%$ of a monolayer (ML) and two representative structures for lower coverages, namely the p$(3\times3)$ and the c$(6\times6)$ structures, corresponding to $\theta=1/9\approx11.1\%$ ML and $\theta=1/18\approx5.6\%$ ML, respectively. For each coverage, Na atoms are adsorbed in a fourfold hollow site. Slabs with 24 layers of copper atoms have been employed, based on a $(12\times12)$ multiple of the primitive Cu(100) surface unit cell with fully-relaxed interlayer distances and a surface lattice constant $a_{100} = 2.556~\text{\AA}$ that is commensurate with all three overlayer structures. During the MD simulations the bottommost layer has conveniently been kept frozen without any consequences for the reported lifetimes, which have all been thoroughly checked for convergence with respect to the lateral cell size of the supercell and the slab thickness \cite{SM}.

The lifetimes obtained for the three overlayer structures when the surface atoms are initially at rest ($T=0$ K) are presented in Fig. \ref{fig:effect_of_coverage} (d), where we compare them to the lifetimes calculated with the ECM. While the ECM predicts accurate lifetimes at the highest coverages, it considerably overestimates the lifetime for the c$(6\times6)$ structure ($\theta=1/18$). Since the parameters of the ECM have been consistently computed with the EAM potential employed in the NEMD simulations, the observed discrepancy has to arise from the breakdown of some of the approximations on which the ECM is based. On the one hand, the continuum approximation for the substrate does not necessarily apply in the context of damping of atomic vibrations. On the other hand, the harmonic approximation that is inherent to the ECM could be too severe to realistically describe the adsorbate coupling to the substrate or the energy transport within the substrate.

We can assess the validity of this harmonic approximation at $T=0$ K by performing a second type of dynamical calculations. The initial excitation of the FT localized in the Na adlayer is described by a superposition of many phonon modes of the strongly-coupled adlayer-substrate system. Our phonon mode projection scheme directly yields the initial components of this phonon wave packet with their amplitudes and phases. The group velocity of this wave packet characterizes the energy transport away from the adlayer into the surface. Unfortunately, this group velocity cannot be easily obtained from the surface phonon band structure, since the wave packet is a linear combination of different (surface) phonon bands at a single phonon wave vector $\boldsymbol{q} = \bar{\Gamma}$. Instead, in order to describe a harmonic bath with harmonic coupling to the adsorbate layer, we simply switch off all the phonon-phonon couplings between the components of this wave packet by analytically propagating them in the eigenbasis defined by the phonon modes, using the aforementioned amplitudes and phases as initial conditions \cite{SM}. Lifetimes $\tau_{\mathrm{harm}}$ calculated from these purely harmonic dynamics, also presented in Fig.~\ref{fig:effect_of_coverage} (d), are in overall good agreement with the values extracted from the NEMD calculations, demonstrating a negligible role of anharmonicity (i.e. phonon-phonon coupling) at $T=0$~K. 

Consequently, we have clearly identified the continuum approximation in the ECM as responsible for the large overestimation of the lifetime at low coverages (cf $\theta=1/18$ in Fig.~\ref{fig:effect_of_coverage}). Vibrationally excited periodic adlayers can transfer energy to the underlying surface only through substrate phonons that are commensurate to the overlayer structure \cite{Lewis-JCP-1998, Pykhtin-PRL-1998}. The range of accessible wavevectors in the SBZ of the substrate becomes narrower for increasing coverages, so that at the highest coverages the adlayer modes can only couple to the substrate phonons with the longest wavelengths. It is in this limit where the elastic continuum approximation is best justified. On the contrary, the lower the coverage, the more phonon modes are backfolded into the smaller and smaller SBZ of the adlayer and thus become accessible -- rationalizing the strong effect of the exact surface phonon band structure as captured by our NEMD simulations.

\begin{figure}
\includegraphics[width=8.6cm]{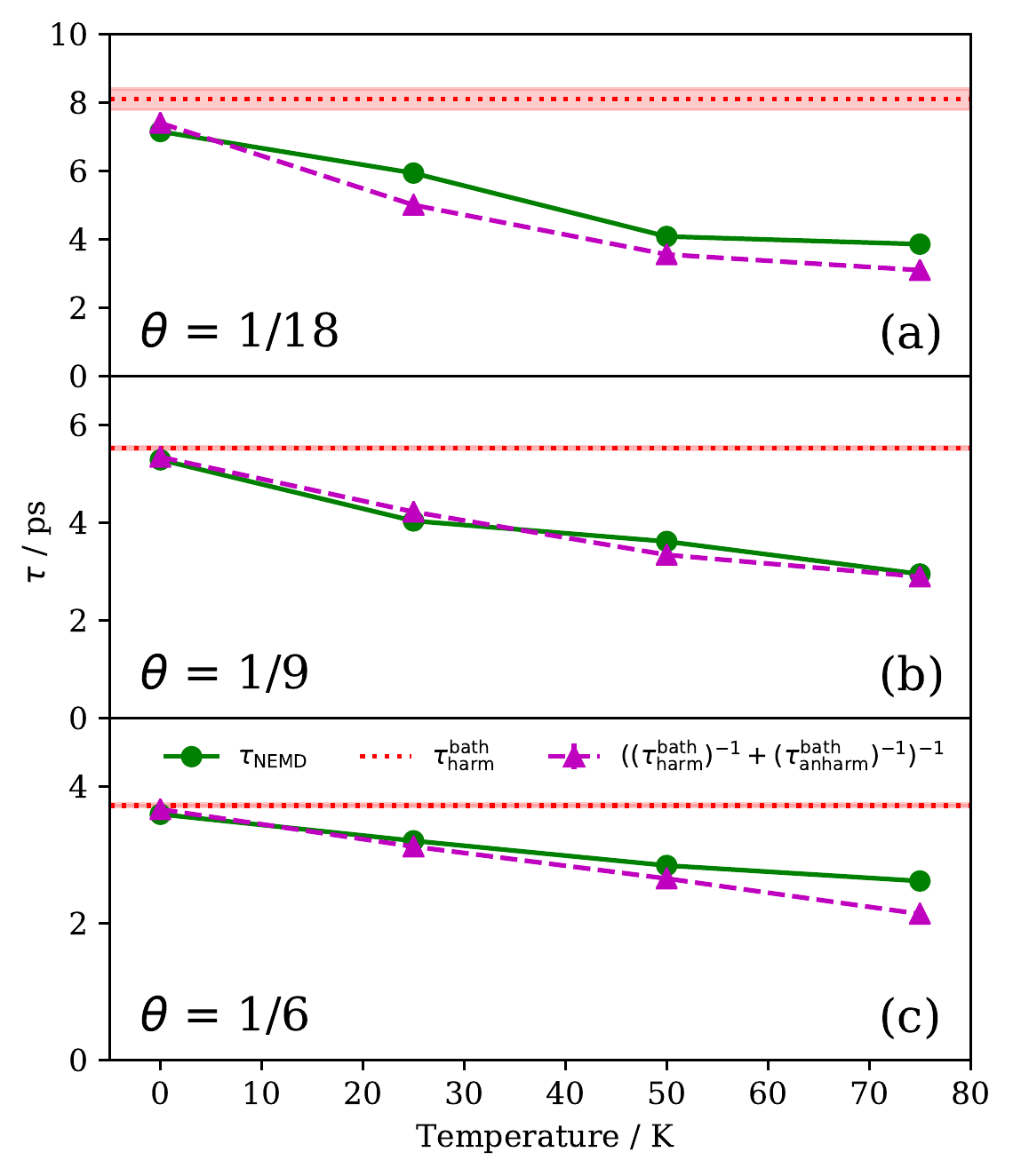}
\caption{\label{fig:effect_of_temperature}%
(color online). Na FT-mode lifetimes as a function of the temperature for the various overlayer structures (panels (a) to (c)). Lifetimes obtained through NEMD calculations (green symbols, solid line to guide the eyes) are compared to the lifetimes extracted from harmonic dynamics (red dotted line) and to the lifetimes estimated from the sum of the harmonic and the anharmonic contribution (violet symbols, dashed line to guide the eyes). Error bars, which are shown as a red shaded area for the harmonic dynamics, reflect the quality of the fits employed to estimate the lifetimes.}
\end{figure}

At finite temperature $T>0$~K the situation is quite different, since we have found strong deviations from the harmonic limit. This is illustrated in Fig. \ref{fig:effect_of_temperature}, where we present Na FT-mode lifetimes calculated in a temperature range at which Na atom diffusion is still negligible ($T<100$ K). For all coverages we observe a considerable reduction of $\tau_{\mathrm{NEMD}}$ at $T=75$~K, which even amounts to about 50\% at $\theta = 1/18$. This temperature dependence can be rationalized by a simple model that separates a temperature-independent harmonic contribution $\tau_{\mathrm{harm}}$ and temperature-dependent anharmonic contribution $\tau_{\mathrm{anharm}}$ to the lifetime. We use $\tau_{\mathrm{harm}}$ as described above for the contribution resulting from the (analytic) propagation of the initially excited phonon wavepacket in the harmonic system defined by the phonon modes. In order to quantify the anharmonic contribution $\tau_{\mathrm{anharm}}$, we change the initial conditions in our NEMD calculations to initially exciting the single phonon eigenmode at $\bar{\Gamma}$ that has the largest overlap with the FT localized in the Na adlayer excited otherwise \cite{SM}. The initial excitation energy in this mode equilibrates because of phonon-phonon coupling in the anharmonic bath with anharmonic coupling to the adsorbate layer. Assuming that the harmonic and anharmonic contributions to the lifetime are separable in this way, the corresponding rates should be additive:
\begin{equation}
\frac{1}{\tau_{\mathrm{NEMD}}} = \frac{1}{\tau_{\mathrm{harm}}} + \frac{1}{\tau_{\mathrm{anharm}}},
\label{eq:model_ph}
\end{equation}
Using our calculated results for $\tau_{\mathrm{harm}}$ and $\tau_{\mathrm{anharm}}$, Fig.~\ref{fig:effect_of_temperature} shows that this simple model indeed almost quantitatively reproduce the temperature dependence of  $\tau_{\mathrm{NEMD}}$ for all coverages. Consistent with our results above, this model also confirms the negligible role of anharmonicity at $T=0$~K, i.e. $\tau_{\mathrm{NEMD}} \approx \tau_{\mathrm{harm}}$ since $ \tau_{\mathrm{anharm}} > 80~\text{ps} \gg \tau_{\mathrm{harm}}$. For increasing temperatures, $\tau_{\mathrm{anharm}}$ rapidly decreases ($\tau_{\mathrm{anharm}}\approx5$ ps at 75 K) and we consistently observe increasing deviations from the harmonic limit. As expected, the temperature dependence of $\tau_{\mathrm{NEMD}}$ is therefore completely given by the anharmonic contribution which quantifies the increasing relevance of phonon-phonon couplings for the lifetime with increasing temperatures.

\begin{figure}
\includegraphics[width=8.6cm]{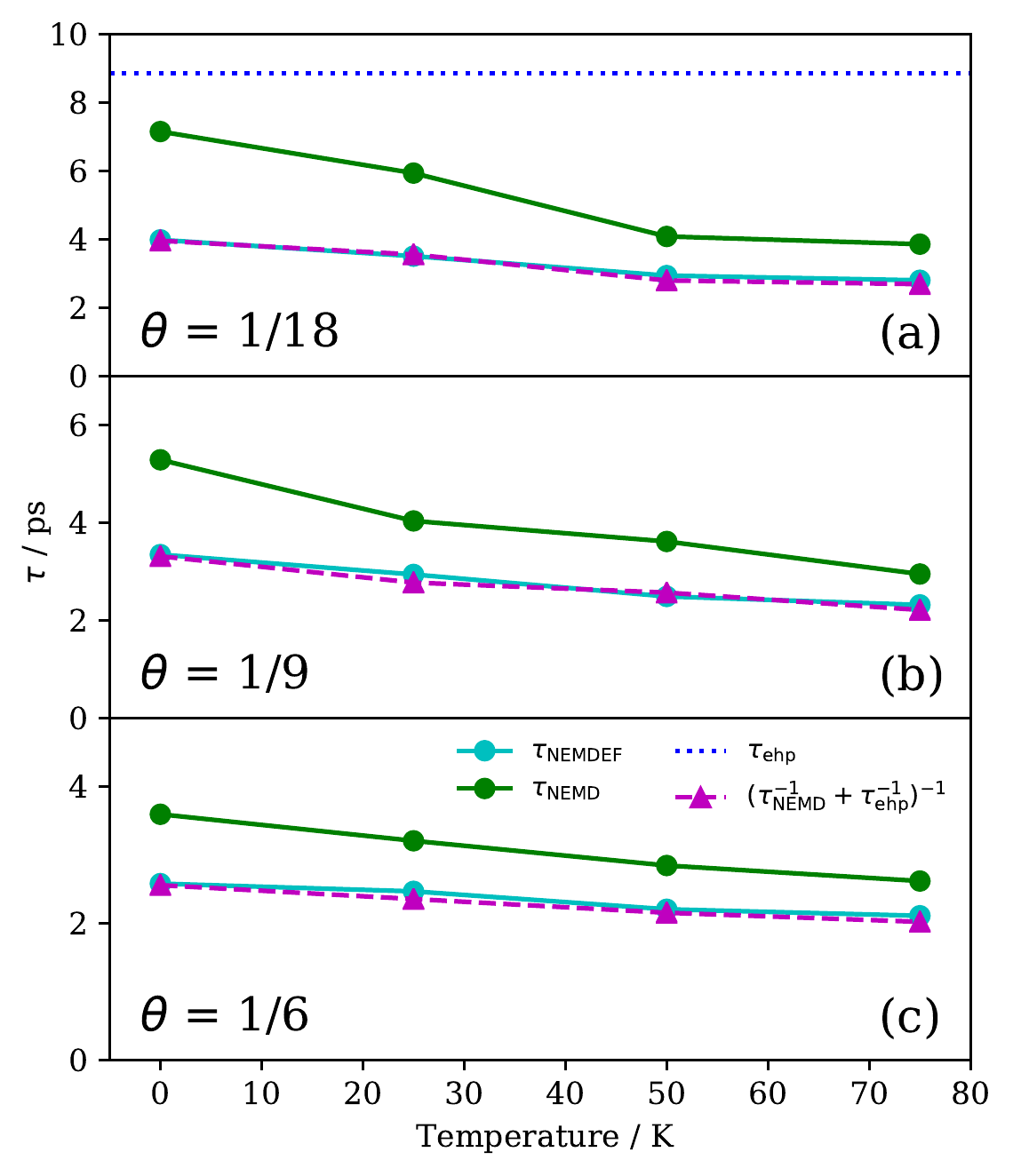}
\caption{\label{fig:effect_of_ehp}%
(color online). Same as Fig. \ref{fig:effect_of_temperature}, but the 
lifetimes obtained through NEMD calculations (green symbols, solid line to guide the eyes)
are here compared to the lifetimes obtained from NEMDEF calculations (cyan symbols,
 solid line to guide the eyes) and to 
the lifetimes estimated from the sum of the phononic and the electronic contribution 
(violet symbols, dashed line to guide the eyes). 
The lifetime corresponding to the electronic damping alone is shown as a blue dotted line in (a).
}
\end{figure}

Finally, we investigate the contribution of electron-phonon coupling for the lifetimes by implementing Langevin dynamics on top of our NEMD calculations. The strength of the electron-phonon coupling is given by a (constant) electronic friction (EF) coefficient \cite{HeadGordon-JCP-1995}, for which we choose a value $\eta_{\mathrm{ehp}}$ that has been obtained from an atoms-in-jellium model and which is representative for Na on copper surfaces \cite{Rittmeyer-PRL-2016,SM}.
Results for the lifetime $\tau_{\mathrm{NEMDEF}}$ obtained from these NEMDEF simulations are presented in Fig. \ref{fig:effect_of_ehp}. As expected, $\tau_{\mathrm{NEMDEF}} < \tau_{\mathrm{NEMD}}$, because an additional energy dissipation channel is being accounted for. 


Using the same argument as in Eq.~\ref{eq:model_ph}, the following equation should hold if the phononic and electronic channels are independent from one antoher:
\begin{equation}
\frac{1}{\tau_{\mathrm{NEMDEF}}} = \frac{1}{\tau_{\mathrm{NEMD}}} + \frac{1}{\tau_{\mathrm{ehp}}},
\label{eq:model_ehp}
\end{equation}
We quantify the eletron-hole pair contribution to the lifetime simply by $\tau_{\mathrm{ehp}} = M/\eta_{\mathrm{ehp}}$, where $M=22.99 \mathrm{u}$ is the adsorbate mass. 
Fig. \ref{fig:effect_of_ehp} shows 
The excellent match between the lifetimes predicted through NEMDEF simulations and the lifetimes calculated from the independent phononic and electronic contributions confirms the additivity of the two damping mechanisms over the whole temperature range considered (Fig. \ref{fig:effect_of_ehp}). 
This conclusion does not depend on the value of the friction coefficient applied, as similar results are obtained through NEMDEF calculations based on a different $\eta_{\mathrm{ehp}}$ estimate \cite{Rittmeyer-PRL-2016, SM}. Note that our comprehensive model predicts the electronic dissipation contribution to represent between 16\% and 23\% of the total damping for $T =75$ K and $\theta \approx 5.6 \%$ , depending on the model employed to estimate $\eta_{\mathrm{ehp}}$ \cite{SM}. This range is in very good agreement with what calculated by Rimmeyer \textit{et al.}, who have estimated a 16\%-26\% range for the relative contribution of the electron-hole pair dissipation channel for Na atoms diffusing on Cu(111) under similar conditions of coverage ($\theta$ = 2.5\% ML) and temperature ($T = 150$ K) \cite{Rittmeyer-PRL-2016}.

%

Overall, our estimate of the Na FT-mode lifetime is not so sensitive on the exact value of the friction coefficient applied, with the largest deviations between the two models being $\approx 20 \%$. However, the lifetime strongly depends on the coverage structure simulated. Due to the absence of a detailed knowledge about the Na overlayer structure in the available measurements \cite{Graham-PRL-1997, Politano-SurfSciRep-2013}, we cannot perform a straightforward theory-experiment comparison using these data. We note, however, that the predicted temperature dependence is consistent with experimental observations of significant lifetime decrease with increasing temperature \cite{Graham-PRL-1997, Politano-SurfSciRep-2013}. Lifetime measurements for well-characterized overlayer structures would be desirable for further testing and developments of theoretical models that aim at describing the energy exchange at interphases. 
\begin{acknowledgments}
J.M. is grateful for financial support from the Netherlands Organisation for Scientific Research (NWO) under VIDI Grant No. 723.014.009.
\end{acknowledgments}
\bibliography{Manuscript}

\begin{thebibliography}{31}%
\makeatletter
\providecommand \@ifxundefined [1]{%
 \@ifx{#1\undefined}
}%
\providecommand \@ifnum [1]{%
 \ifnum #1\expandafter \@firstoftwo
 \else \expandafter \@secondoftwo
 \fi
}%
\providecommand \@ifx [1]{%
 \ifx #1\expandafter \@firstoftwo
 \else \expandafter \@secondoftwo
 \fi
}%
\providecommand \natexlab [1]{#1}%
\providecommand \enquote  [1]{``#1''}%
\providecommand \bibnamefont  [1]{#1}%
\providecommand \bibfnamefont [1]{#1}%
\providecommand \citenamefont [1]{#1}%
\providecommand \href@noop [0]{\@secondoftwo}%
\providecommand \href [0]{\begingroup \@sanitize@url \@href}%
\providecommand \@href[1]{\@@startlink{#1}\@@href}%
\providecommand \@@href[1]{\endgroup#1\@@endlink}%
\providecommand \@sanitize@url [0]{\catcode `\\12\catcode `\$12\catcode
  `\&12\catcode `\#12\catcode `\^12\catcode `\_12\catcode `\%12\relax}%
\providecommand \@@startlink[1]{}%
\providecommand \@@endlink[0]{}%
\providecommand \url  [0]{\begingroup\@sanitize@url \@url }%
\providecommand \@url [1]{\endgroup\@href {#1}{\urlprefix }}%
\providecommand \urlprefix  [0]{URL }%
\providecommand \Eprint [0]{\href }%
\providecommand \doibase [0]{http://dx.doi.org/}%
\providecommand \selectlanguage [0]{\@gobble}%
\providecommand \bibinfo  [0]{\@secondoftwo}%
\providecommand \bibfield  [0]{\@secondoftwo}%
\providecommand \translation [1]{[#1]}%
\providecommand \BibitemOpen [0]{}%
\providecommand \bibitemStop [0]{}%
\providecommand \bibitemNoStop [0]{.\EOS\space}%
\providecommand \EOS [0]{\spacefactor3000\relax}%
\providecommand \BibitemShut  [1]{\csname bibitem#1\endcsname}%
\let\auto@bib@innerbib\@empty
\bibitem [{\citenamefont {Somorjai}\ and\ \citenamefont
  {Li}(2011)}]{Somorjai-PNAS-2011}%
  \BibitemOpen
  \bibfield  {author} {\bibinfo {author} {\bibfnamefont {G.~A.}\ \bibnamefont
  {Somorjai}}\ and\ \bibinfo {author} {\bibfnamefont {Y.}~\bibnamefont {Li}},\
  }\href {\doibase 10.1073/pnas.1006669107} {\bibfield  {journal} {\bibinfo
  {journal} {Proc. Natl. Acad. Sci. U. S. A.}\ }\textbf {\bibinfo {volume}
  {108}},\ \bibinfo {pages} {917} (\bibinfo {year} {2011})}\BibitemShut
  {NoStop}%
\bibitem [{\citenamefont {Nilsson}\ \emph {et~al.}(2008)\citenamefont
  {Nilsson}, \citenamefont {Pettersson},\ and\ \citenamefont
  {N\o{}rskov~(Eds.)}}]{Nilsson-ChemicalBondingAtSurfacesAndInterfaces-2008}%
  \BibitemOpen
  \bibfield  {author} {\bibinfo {author} {\bibfnamefont {A.}~\bibnamefont
  {Nilsson}}, \bibinfo {author} {\bibfnamefont {L.~G.~M.}\ \bibnamefont
  {Pettersson}}, \ and\ \bibinfo {author} {\bibfnamefont {J.~K.}\ \bibnamefont
  {N\o{}rskov~(Eds.)}},\ }\href@noop {} {\emph {\bibinfo {title} {{Chemical
  Bonding at Surfaces and Interfaces}}}}\ (\bibinfo  {publisher} {Elsevier},\
  \bibinfo {address} {Amsterdam},\ \bibinfo {year} {2008})\BibitemShut
  {NoStop}%
\bibitem [{\citenamefont {Graham}\ \emph
  {et~al.}(1997{\natexlab{a}})\citenamefont {Graham}, \citenamefont {Hofmann},
  \citenamefont {Toennies}, \citenamefont {Chen},\ and\ \citenamefont
  {Ying}}]{Graham-PRL-1997}%
  \BibitemOpen
  \bibfield  {author} {\bibinfo {author} {\bibfnamefont {A.~P.}\ \bibnamefont
  {Graham}}, \bibinfo {author} {\bibfnamefont {F.}~\bibnamefont {Hofmann}},
  \bibinfo {author} {\bibfnamefont {J.~P.}\ \bibnamefont {Toennies}}, \bibinfo
  {author} {\bibfnamefont {L.~Y.}\ \bibnamefont {Chen}}, \ and\ \bibinfo
  {author} {\bibfnamefont {S.~C.}\ \bibnamefont {Ying}},\ }\href {\doibase
  10.1103/PhysRevLett.78.3900} {\bibfield  {journal} {\bibinfo  {journal}
  {Phys. Rev. Lett.}\ }\textbf {\bibinfo {volume} {78}},\ \bibinfo {pages}
  {3900} (\bibinfo {year} {1997}{\natexlab{a}})}\BibitemShut {NoStop}%
\bibitem [{\citenamefont {Graham}(2003)}]{Graham-SurfSciRep-2003}%
  \BibitemOpen
  \bibfield  {author} {\bibinfo {author} {\bibfnamefont {A.~P.}\ \bibnamefont
  {Graham}},\ }\href {\doibase 10.1016/S0167-5729(03)00012-8} {\bibfield
  {journal} {\bibinfo  {journal} {Surf. Sci. Rep.}\ }\textbf {\bibinfo {volume}
  {49}},\ \bibinfo {pages} {115} (\bibinfo {year} {2003})}\BibitemShut
  {NoStop}%
\bibitem [{\citenamefont {Politano}\ \emph {et~al.}(2013)\citenamefont
  {Politano}, \citenamefont {Chiarello}, \citenamefont {Benedek}, \citenamefont
  {Chulkov},\ and\ \citenamefont {Echenique}}]{Politano-SurfSciRep-2013}%
  \BibitemOpen
  \bibfield  {author} {\bibinfo {author} {\bibfnamefont {A.}~\bibnamefont
  {Politano}}, \bibinfo {author} {\bibfnamefont {G.}~\bibnamefont {Chiarello}},
  \bibinfo {author} {\bibfnamefont {G.}~\bibnamefont {Benedek}}, \bibinfo
  {author} {\bibfnamefont {E.}~\bibnamefont {Chulkov}}, \ and\ \bibinfo
  {author} {\bibfnamefont {P.}~\bibnamefont {Echenique}},\ }\href {\doibase
  10.1016/j.surfrep.2013.07.001} {\bibfield  {journal} {\bibinfo  {journal}
  {Surf. Sci. Rep.}\ }\textbf {\bibinfo {volume} {68}},\ \bibinfo {pages} {305}
  (\bibinfo {year} {2013})}\BibitemShut {NoStop}%
\bibitem [{\citenamefont {Rittmeyer}\ \emph {et~al.}(2016)\citenamefont
  {Rittmeyer}, \citenamefont {Ward}, \citenamefont {G{\"{u}}tlein},
  \citenamefont {Ellis}, \citenamefont {Allison},\ and\ \citenamefont
  {Reuter}}]{Rittmeyer-PRL-2016}%
  \BibitemOpen
  \bibfield  {author} {\bibinfo {author} {\bibfnamefont {S.~P.}\ \bibnamefont
  {Rittmeyer}}, \bibinfo {author} {\bibfnamefont {D.~J.}\ \bibnamefont {Ward}},
  \bibinfo {author} {\bibfnamefont {P.}~\bibnamefont {G{\"{u}}tlein}}, \bibinfo
  {author} {\bibfnamefont {J.}~\bibnamefont {Ellis}}, \bibinfo {author}
  {\bibfnamefont {W.}~\bibnamefont {Allison}}, \ and\ \bibinfo {author}
  {\bibfnamefont {K.}~\bibnamefont {Reuter}},\ }\href {\doibase
  10.1103/PhysRevLett.117.196001} {\bibfield  {journal} {\bibinfo  {journal}
  {Phys. Rev. Lett.}\ }\textbf {\bibinfo {volume} {117}},\ \bibinfo {pages}
  {196001} (\bibinfo {year} {2016})}\BibitemShut {NoStop}%
\bibitem [{\citenamefont {Juaristi}\ \emph {et~al.}(2008)\citenamefont
  {Juaristi}, \citenamefont {Alducin}, \citenamefont {D{\'i}ez~Mui{\~n}o},
  \citenamefont {Busnengo},\ and\ \citenamefont {Salin}}]{Juaristi-PRL-2008}%
  \BibitemOpen
  \bibfield  {author} {\bibinfo {author} {\bibfnamefont {J.~I.}\ \bibnamefont
  {Juaristi}}, \bibinfo {author} {\bibfnamefont {M.}~\bibnamefont {Alducin}},
  \bibinfo {author} {\bibfnamefont {R.}~\bibnamefont {D{\'i}ez~Mui{\~n}o}},
  \bibinfo {author} {\bibfnamefont {H.~F.}\ \bibnamefont {Busnengo}}, \ and\
  \bibinfo {author} {\bibfnamefont {A.}~\bibnamefont {Salin}},\ }\href
  {\doibase 10.1103/PhysRevLett.100.116102} {\bibfield  {journal} {\bibinfo
  {journal} {Phys. Rev. Lett.}\ }\textbf {\bibinfo {volume} {100}},\ \bibinfo
  {pages} {116102} (\bibinfo {year} {2008})}\BibitemShut {NoStop}%
\bibitem [{\citenamefont {Blanco-Rey}\ \emph {et~al.}(2014)\citenamefont
  {Blanco-Rey}, \citenamefont {Juaristi}, \citenamefont {{D{\'{i}}ez
  Mui{\~{n}}o}}, \citenamefont {Busnengo}, \citenamefont {Kroes},\ and\
  \citenamefont {Alducin}}]{BlancoRey-PRL-2014}%
  \BibitemOpen
  \bibfield  {author} {\bibinfo {author} {\bibfnamefont {M.}~\bibnamefont
  {Blanco-Rey}}, \bibinfo {author} {\bibfnamefont {J.~I.}\ \bibnamefont
  {Juaristi}}, \bibinfo {author} {\bibfnamefont {R.}~\bibnamefont {{D{\'{i}}ez
  Mui{\~{n}}o}}}, \bibinfo {author} {\bibfnamefont {H.~F.}\ \bibnamefont
  {Busnengo}}, \bibinfo {author} {\bibfnamefont {G.~J.}\ \bibnamefont {Kroes}},
  \ and\ \bibinfo {author} {\bibfnamefont {M.}~\bibnamefont {Alducin}},\ }\href
  {\doibase 10.1103/PhysRevLett.112.103203} {\bibfield  {journal} {\bibinfo
  {journal} {Phys. Rev. Lett.}\ }\textbf {\bibinfo {volume} {112}},\ \bibinfo
  {pages} {103203} (\bibinfo {year} {2014})}\BibitemShut {NoStop}%
\bibitem [{\citenamefont {Rittmeyer}\ \emph {et~al.}(2015)\citenamefont
  {Rittmeyer}, \citenamefont {Meyer}, \citenamefont {Juaristi},\ and\
  \citenamefont {Reuter}}]{Rittmeyer-PRL-2015}%
  \BibitemOpen
  \bibfield  {author} {\bibinfo {author} {\bibfnamefont {S.~P.}\ \bibnamefont
  {Rittmeyer}}, \bibinfo {author} {\bibfnamefont {J.}~\bibnamefont {Meyer}},
  \bibinfo {author} {\bibfnamefont {J.~I.}\ \bibnamefont {Juaristi}}, \ and\
  \bibinfo {author} {\bibfnamefont {K.}~\bibnamefont {Reuter}},\ }\href
  {\doibase 10.1103/PhysRevLett.115.046102} {\bibfield  {journal} {\bibinfo
  {journal} {Phys. Rev. Lett.}\ }\textbf {\bibinfo {volume} {115}},\ \bibinfo
  {pages} {046102} (\bibinfo {year} {2015})}\BibitemShut {NoStop}%
\bibitem [{\citenamefont {Tully}\ \emph {et~al.}(1993)\citenamefont {Tully},
  \citenamefont {Gomez},\ and\ \citenamefont
  {Head-Gordon}}]{Tully-JVacSciTechnol-1993}%
  \BibitemOpen
  \bibfield  {author} {\bibinfo {author} {\bibfnamefont {J.~C.}\ \bibnamefont
  {Tully}}, \bibinfo {author} {\bibfnamefont {M.}~\bibnamefont {Gomez}}, \ and\
  \bibinfo {author} {\bibfnamefont {M.}~\bibnamefont {Head-Gordon}},\ }\href
  {\doibase 10.1116/1.578522} {\bibfield  {journal} {\bibinfo  {journal} {J.
  Vac. Sci. Technol. A}\ }\textbf {\bibinfo {volume} {11}},\ \bibinfo {pages}
  {1914} (\bibinfo {year} {1993})}\BibitemShut {NoStop}%
\bibitem [{\citenamefont {Tully}(1980)}]{Tully-JCP-1980}%
  \BibitemOpen
  \bibfield  {author} {\bibinfo {author} {\bibfnamefont {J.~C.}\ \bibnamefont
  {Tully}},\ }\href {\doibase 10.1063/1.440287} {\bibfield  {journal} {\bibinfo
   {journal} {J. Chem. Phys.}\ }\textbf {\bibinfo {volume} {73}},\ \bibinfo
  {pages} {1975} (\bibinfo {year} {1980})}\BibitemShut {NoStop}%
\bibitem [{\citenamefont {{Head-Gordon}}\ and\ \citenamefont
  {Tully}(1995)}]{HeadGordon-JCP-1995}%
  \BibitemOpen
  \bibfield  {author} {\bibinfo {author} {\bibfnamefont {M.}~\bibnamefont
  {{Head-Gordon}}}\ and\ \bibinfo {author} {\bibfnamefont {J.~C.}\ \bibnamefont
  {Tully}},\ }\href {\doibase 10.1063/1.469915} {\bibfield  {journal} {\bibinfo
   {journal} {J. Chem. Phys.}\ }\textbf {\bibinfo {volume} {103}},\ \bibinfo
  {pages} {10137} (\bibinfo {year} {1995})}\BibitemShut {NoStop}%
\bibitem [{\citenamefont {Chen}\ \emph {et~al.}(2018)\citenamefont {Chen},
  \citenamefont {Lau}, \citenamefont {Schwarzer}, \citenamefont {Meyer},
  \citenamefont {Verma},\ and\ \citenamefont {Wodtke}}]{Chen-Science-2018}%
  \BibitemOpen
  \bibfield  {author} {\bibinfo {author} {\bibfnamefont {L.}~\bibnamefont
  {Chen}}, \bibinfo {author} {\bibfnamefont {J.~A.}\ \bibnamefont {Lau}},
  \bibinfo {author} {\bibfnamefont {D.}~\bibnamefont {Schwarzer}}, \bibinfo
  {author} {\bibfnamefont {J.}~\bibnamefont {Meyer}}, \bibinfo {author}
  {\bibfnamefont {V.~B.}\ \bibnamefont {Verma}}, \ and\ \bibinfo {author}
  {\bibfnamefont {A.~M.}\ \bibnamefont {Wodtke}},\ }\href@noop {} {\bibfield
  {journal} {\bibinfo  {journal} {accepted by Science}\ } (\bibinfo {year}
  {2018})}\BibitemShut {NoStop}%
\bibitem [{\citenamefont {Maradudin}\ and\ \citenamefont
  {Fein}(1962)}]{Maradudin-PhysRev-1962}%
  \BibitemOpen
  \bibfield  {author} {\bibinfo {author} {\bibfnamefont {A.~A.}\ \bibnamefont
  {Maradudin}}\ and\ \bibinfo {author} {\bibfnamefont {A.~E.}\ \bibnamefont
  {Fein}},\ }\href {\doibase 10.1103/PhysRev.128.2589} {\bibfield  {journal}
  {\bibinfo  {journal} {Phys. Rev.}\ }\textbf {\bibinfo {volume} {128}},\
  \bibinfo {pages} {2589} (\bibinfo {year} {1962})}\BibitemShut {NoStop}%
\bibitem [{\citenamefont {Meyer}(2011)}]{Meyer-PhDThesis}%
  \BibitemOpen
  \bibfield  {author} {\bibinfo {author} {\bibfnamefont {J.}~\bibnamefont
  {Meyer}},\ }\emph {\bibinfo {title} {Ab initio Modeling of Energy Dissipation
  during Chemical Reactions at Transition Metal Surfaces}},\ \href@noop {}
  {Ph.D. thesis},\ \bibinfo  {school} {Freie Universit\"at Berlin} (\bibinfo
  {year} {2011})\BibitemShut {NoStop}%
\bibitem [{\citenamefont {Meyer}\ and\ \citenamefont
  {Reuter}(2014)}]{Meyer-Angew-2014}%
  \BibitemOpen
  \bibfield  {author} {\bibinfo {author} {\bibfnamefont {J.}~\bibnamefont
  {Meyer}}\ and\ \bibinfo {author} {\bibfnamefont {K.}~\bibnamefont {Reuter}},\
  }\href {\doibase 10.1002/anie.201400066} {\bibfield  {journal} {\bibinfo
  {journal} {Angew. Chem. Int. Ed.}\ }\textbf {\bibinfo {volume} {53}},\
  \bibinfo {pages} {4721} (\bibinfo {year} {2014})}\BibitemShut {NoStop}%
\bibitem [{\citenamefont {Persson}\ and\ \citenamefont
  {Ryberg}(1985)}]{Persson-PRB-1985}%
  \BibitemOpen
  \bibfield  {author} {\bibinfo {author} {\bibfnamefont {B.~N.~J.}\
  \bibnamefont {Persson}}\ and\ \bibinfo {author} {\bibfnamefont
  {R.}~\bibnamefont {Ryberg}},\ }\href {\doibase 10.1103/PhysRevB.32.3586}
  {\bibfield  {journal} {\bibinfo  {journal} {Phys. Rev. B}\ }\textbf {\bibinfo
  {volume} {32}},\ \bibinfo {pages} {3586} (\bibinfo {year}
  {1985})}\BibitemShut {NoStop}%
\bibitem [{\citenamefont {Lewis}\ \emph {et~al.}(1998)\citenamefont {Lewis},
  \citenamefont {Pykhtin}, \citenamefont {Mele},\ and\ \citenamefont
  {Rappe}}]{Lewis-JCP-1998}%
  \BibitemOpen
  \bibfield  {author} {\bibinfo {author} {\bibfnamefont {S.~P.}\ \bibnamefont
  {Lewis}}, \bibinfo {author} {\bibfnamefont {M.~V.}\ \bibnamefont {Pykhtin}},
  \bibinfo {author} {\bibfnamefont {E.~J.}\ \bibnamefont {Mele}}, \ and\
  \bibinfo {author} {\bibfnamefont {A.~M.}\ \bibnamefont {Rappe}},\ }\href
  {\doibase 10.1063/1.475478} {\bibfield  {journal} {\bibinfo  {journal} {J.
  Chem. Phys.}\ }\textbf {\bibinfo {volume} {108}},\ \bibinfo {pages} {1157}
  (\bibinfo {year} {1998})}\BibitemShut {NoStop}%
\bibitem [{\citenamefont {Pykhtin}\ \emph {et~al.}(1998)\citenamefont
  {Pykhtin}, \citenamefont {Lewis}, \citenamefont {Mele},\ and\ \citenamefont
  {Rappe}}]{Pykhtin-PRL-1998}%
  \BibitemOpen
  \bibfield  {author} {\bibinfo {author} {\bibfnamefont {M.~V.}\ \bibnamefont
  {Pykhtin}}, \bibinfo {author} {\bibfnamefont {S.~P.}\ \bibnamefont {Lewis}},
  \bibinfo {author} {\bibfnamefont {E.~J.}\ \bibnamefont {Mele}}, \ and\
  \bibinfo {author} {\bibfnamefont {A.~M.}\ \bibnamefont {Rappe}},\ }\href
  {\doibase 10.1103/PhysRevLett.81.5940} {\bibfield  {journal} {\bibinfo
  {journal} {Phys. Rev. Lett.}\ }\textbf {\bibinfo {volume} {81}},\ \bibinfo
  {pages} {5940} (\bibinfo {year} {1998})}\BibitemShut {NoStop}%
\bibitem [{\citenamefont {Lackey}\ and\ \citenamefont
  {King}(1987)}]{Lackey-JCSFT1-1987}%
  \BibitemOpen
  \bibfield  {author} {\bibinfo {author} {\bibfnamefont {D.}~\bibnamefont
  {Lackey}}\ and\ \bibinfo {author} {\bibfnamefont {D.~A.}\ \bibnamefont
  {King}},\ }\href {\doibase 10.1039/F19878302001} {\bibfield  {journal}
  {\bibinfo  {journal} {J. Chem. Soc.{,} Faraday Trans. 1}\ }\textbf {\bibinfo
  {volume} {83}},\ \bibinfo {pages} {2001} (\bibinfo {year}
  {1987})}\BibitemShut {NoStop}%
\bibitem [{\citenamefont {Daw}\ and\ \citenamefont
  {Baskes}(1983)}]{Daw-PRL-1983}%
  \BibitemOpen
  \bibfield  {author} {\bibinfo {author} {\bibfnamefont {M.~S.}\ \bibnamefont
  {Daw}}\ and\ \bibinfo {author} {\bibfnamefont {M.~I.}\ \bibnamefont
  {Baskes}},\ }\href {\doibase 10.1103/PhysRevLett.50.1285} {\bibfield
  {journal} {\bibinfo  {journal} {Phys. Rev. Lett.}\ }\textbf {\bibinfo
  {volume} {50}},\ \bibinfo {pages} {1285} (\bibinfo {year}
  {1983})}\BibitemShut {NoStop}%
\bibitem [{\citenamefont {Senet}\ \emph {et~al.}(1999)\citenamefont {Senet},
  \citenamefont {Toennies},\ and\ \citenamefont
  {Witte}}]{Senet-ChemPhysLett-1999}%
  \BibitemOpen
  \bibfield  {author} {\bibinfo {author} {\bibfnamefont {P.}~\bibnamefont
  {Senet}}, \bibinfo {author} {\bibfnamefont {J.}~\bibnamefont {Toennies}}, \
  and\ \bibinfo {author} {\bibfnamefont {G.}~\bibnamefont {Witte}},\ }\href
  {\doibase 10.1016/S0009-2614(98)01314-1} {\bibfield  {journal} {\bibinfo
  {journal} {Chem. Phys. Lett.}\ }\textbf {\bibinfo {volume} {299}},\ \bibinfo
  {pages} {389} (\bibinfo {year} {1999})}\BibitemShut {NoStop}%
\bibitem [{SM()}]{SM}%
  \BibitemOpen
  \href@noop {} {}\bibinfo {note} {See Supplemental Material at
  LINK.}\BibitemShut {Stop}%
\bibitem [{\citenamefont {Graham}\ \emph
  {et~al.}(1997{\natexlab{b}})\citenamefont {Graham}, \citenamefont {Hofmann},
  \citenamefont {Toennies}, \citenamefont {Chen},\ and\ \citenamefont
  {Ying}}]{Graham-PRB-1997-1}%
  \BibitemOpen
  \bibfield  {author} {\bibinfo {author} {\bibfnamefont {A.~P.}\ \bibnamefont
  {Graham}}, \bibinfo {author} {\bibfnamefont {F.}~\bibnamefont {Hofmann}},
  \bibinfo {author} {\bibfnamefont {J.~P.}\ \bibnamefont {Toennies}}, \bibinfo
  {author} {\bibfnamefont {L.~Y.}\ \bibnamefont {Chen}}, \ and\ \bibinfo
  {author} {\bibfnamefont {S.~C.}\ \bibnamefont {Ying}},\ }\href {\doibase
  10.1103/PhysRevB.56.10567} {\bibfield  {journal} {\bibinfo  {journal} {Phys.
  Rev. B}\ }\textbf {\bibinfo {volume} {56}},\ \bibinfo {pages} {10567}
  (\bibinfo {year} {1997}{\natexlab{b}})}\BibitemShut {NoStop}%
\bibitem [{\citenamefont {Plimpton}(1995)}]{Plimpton-JCompPhys-1995}%
  \BibitemOpen
  \bibfield  {author} {\bibinfo {author} {\bibfnamefont {S.}~\bibnamefont
  {Plimpton}},\ }\href {\doibase 10.1006/jcph.1995.1039} {\bibfield  {journal}
  {\bibinfo  {journal} {J. Comp. Phys.}\ }\textbf {\bibinfo {volume} {117}},\
  \bibinfo {pages} {1} (\bibinfo {year} {1995})}\BibitemShut {NoStop}%
\bibitem [{\citenamefont {West}\ and\ \citenamefont
  {Estreicher}(2007)}]{West-PRB-2007}%
  \BibitemOpen
  \bibfield  {author} {\bibinfo {author} {\bibfnamefont {D.}~\bibnamefont
  {West}}\ and\ \bibinfo {author} {\bibfnamefont {S.~K.}\ \bibnamefont
  {Estreicher}},\ }\href {\doibase 10.1103/PhysRevB.75.075206} {\bibfield
  {journal} {\bibinfo  {journal} {Phys. Rev. B}\ }\textbf {\bibinfo {volume}
  {75}},\ \bibinfo {pages} {075206} (\bibinfo {year} {2007})}\BibitemShut
  {NoStop}%
\bibitem [{\citenamefont {Watanabe}\ \emph {et~al.}(2005)\citenamefont
  {Watanabe}, \citenamefont {Takagi},\ and\ \citenamefont
  {Matsumoto}}]{watanabe2005}%
  \BibitemOpen
  \bibfield  {author} {\bibinfo {author} {\bibfnamefont {K.}~\bibnamefont
  {Watanabe}}, \bibinfo {author} {\bibfnamefont {N.}~\bibnamefont {Takagi}}, \
  and\ \bibinfo {author} {\bibfnamefont {Y.}~\bibnamefont {Matsumoto}},\ }\href
  {\doibase 10.1039/B507128C} {\bibfield  {journal} {\bibinfo  {journal} {Phys.
  Chem. Chem. Phys.}\ }\textbf {\bibinfo {volume} {7}},\ \bibinfo {pages}
  {2697} (\bibinfo {year} {2005})}\BibitemShut {NoStop}%
\bibitem [{\citenamefont {Fuyuki}\ \emph {et~al.}(2007)\citenamefont {Fuyuki},
  \citenamefont {Watanabe}, \citenamefont {Ino}, \citenamefont {Petek},\ and\
  \citenamefont {Matsumoto}}]{fuyuki2007}%
  \BibitemOpen
  \bibfield  {author} {\bibinfo {author} {\bibfnamefont {M.}~\bibnamefont
  {Fuyuki}}, \bibinfo {author} {\bibfnamefont {K.}~\bibnamefont {Watanabe}},
  \bibinfo {author} {\bibfnamefont {D.}~\bibnamefont {Ino}}, \bibinfo {author}
  {\bibfnamefont {H.}~\bibnamefont {Petek}}, \ and\ \bibinfo {author}
  {\bibfnamefont {Y.}~\bibnamefont {Matsumoto}},\ }\href {\doibase
  10.1103/PhysRevB.76.115427} {\bibfield  {journal} {\bibinfo  {journal} {Phys.
  Rev. B}\ }\textbf {\bibinfo {volume} {76}},\ \bibinfo {pages} {115427}
  (\bibinfo {year} {2007})}\BibitemShut {NoStop}%
\bibitem [{\citenamefont {McGaughey}\ and\ \citenamefont
  {Kaviany}(2004)}]{McGaughey-PRB-2004}%
  \BibitemOpen
  \bibfield  {author} {\bibinfo {author} {\bibfnamefont {A.~J.~H.}\
  \bibnamefont {McGaughey}}\ and\ \bibinfo {author} {\bibfnamefont
  {M.}~\bibnamefont {Kaviany}},\ }\href {\doibase 10.1103/PhysRevB.69.094303}
  {\bibfield  {journal} {\bibinfo  {journal} {Phys. Rev. B}\ }\textbf {\bibinfo
  {volume} {69}},\ \bibinfo {pages} {1} (\bibinfo {year} {2004})}\BibitemShut
  {NoStop}%
\bibitem [{\citenamefont {West}\ and\ \citenamefont
  {Estreicher}(2006)}]{West-PRL-2006}%
  \BibitemOpen
  \bibfield  {author} {\bibinfo {author} {\bibfnamefont {D.}~\bibnamefont
  {West}}\ and\ \bibinfo {author} {\bibfnamefont {S.~K.}\ \bibnamefont
  {Estreicher}},\ }\href {\doibase 10.1103/PhysRevLett.96.115504} {\bibfield
  {journal} {\bibinfo  {journal} {Phys. Rev. Lett.}\ }\textbf {\bibinfo
  {volume} {96}},\ \bibinfo {pages} {115504} (\bibinfo {year}
  {2006})}\BibitemShut {NoStop}%
\bibitem [{\citenamefont {Graham}\ and\ \citenamefont
  {Toennies}(1997)}]{Graham-PRB-1997-2}%
  \BibitemOpen
  \bibfield  {author} {\bibinfo {author} {\bibfnamefont {A.~P.}\ \bibnamefont
  {Graham}}\ and\ \bibinfo {author} {\bibfnamefont {J.~P.}\ \bibnamefont
  {Toennies}},\ }\href {\doibase 10.1103/PhysRevB.56.15378} {\bibfield
  {journal} {\bibinfo  {journal} {Phys. Rev. B}\ }\textbf {\bibinfo {volume}
  {56}},\ \bibinfo {pages} {15378} (\bibinfo {year} {1997})}\BibitemShut
  {NoStop}%
\end{thebibliography}%


\begin{thebibliography}{47}%
\makeatletter
\providecommand \@ifxundefined [1]{%
 \@ifx{#1\undefined}
}%
\providecommand \@ifnum [1]{%
 \ifnum #1\expandafter \@firstoftwo
 \else \expandafter \@secondoftwo
 \fi
}%
\providecommand \@ifx [1]{%
 \ifx #1\expandafter \@firstoftwo
 \else \expandafter \@secondoftwo
 \fi
}%
\providecommand \natexlab [1]{#1}%
\providecommand \enquote  [1]{``#1''}%
\providecommand \bibnamefont  [1]{#1}%
\providecommand \bibfnamefont [1]{#1}%
\providecommand \citenamefont [1]{#1}%
\providecommand \href@noop [0]{\@secondoftwo}%
\providecommand \href [0]{\begingroup \@sanitize@url \@href}%
\providecommand \@href[1]{\@@startlink{#1}\@@href}%
\providecommand \@@href[1]{\endgroup#1\@@endlink}%
\providecommand \@sanitize@url [0]{\catcode `\\12\catcode `\$12\catcode
  `\&12\catcode `\#12\catcode `\^12\catcode `\_12\catcode `\%12\relax}%
\providecommand \@@startlink[1]{}%
\providecommand \@@endlink[0]{}%
\providecommand \url  [0]{\begingroup\@sanitize@url \@url }%
\providecommand \@url [1]{\endgroup\@href {#1}{\urlprefix }}%
\providecommand \urlprefix  [0]{URL }%
\providecommand \Eprint [0]{\href }%
\providecommand \doibase [0]{http://dx.doi.org/}%
\providecommand \selectlanguage [0]{\@gobble}%
\providecommand \bibinfo  [0]{\@secondoftwo}%
\providecommand \bibfield  [0]{\@secondoftwo}%
\providecommand \translation [1]{[#1]}%
\providecommand \BibitemOpen [0]{}%
\providecommand \bibitemStop [0]{}%
\providecommand \bibitemNoStop [0]{.\EOS\space}%
\providecommand \EOS [0]{\spacefactor3000\relax}%
\providecommand \BibitemShut  [1]{\csname bibitem#1\endcsname}%
\let\auto@bib@innerbib\@empty
\bibitem [{\citenamefont {Daw}\ and\ \citenamefont
  {Baskes}(1983)}]{Daw-PRL-1983}%
  \BibitemOpen
  \bibfield  {author} {\bibinfo {author} {\bibfnamefont {M.~S.}\ \bibnamefont
  {Daw}}\ and\ \bibinfo {author} {\bibfnamefont {M.~I.}\ \bibnamefont
  {Baskes}},\ }\href {\doibase 10.1103/PhysRevLett.50.1285} {\bibfield
  {journal} {\bibinfo  {journal} {Phys. Rev. Lett.}\ }\textbf {\bibinfo
  {volume} {50}},\ \bibinfo {pages} {1285} (\bibinfo {year}
  {1983})}\BibitemShut {NoStop}%
\bibitem [{\citenamefont {Daw}\ and\ \citenamefont
  {Baskes}(1984)}]{Daw-PRB-1984}%
  \BibitemOpen
  \bibfield  {author} {\bibinfo {author} {\bibfnamefont {M.~S.}\ \bibnamefont
  {Daw}}\ and\ \bibinfo {author} {\bibfnamefont {M.~I.}\ \bibnamefont
  {Baskes}},\ }\href {\doibase 10.1103/PhysRevB.29.6443} {\bibfield  {journal}
  {\bibinfo  {journal} {Phys. Rev. B}\ }\textbf {\bibinfo {volume} {29}},\
  \bibinfo {pages} {6443} (\bibinfo {year} {1984})}\BibitemShut {NoStop}%
\bibitem [{\citenamefont {Cleri}\ and\ \citenamefont
  {Rosato}(1993)}]{Cleri-PRB-1993}%
  \BibitemOpen
  \bibfield  {author} {\bibinfo {author} {\bibfnamefont {F.}~\bibnamefont
  {Cleri}}\ and\ \bibinfo {author} {\bibfnamefont {V.}~\bibnamefont {Rosato}},\
  }\href {\doibase 10.1103/PhysRevB.48.22} {\bibfield  {journal} {\bibinfo
  {journal} {Phys. Rev. B}\ }\textbf {\bibinfo {volume} {48}},\ \bibinfo
  {pages} {22} (\bibinfo {year} {1993})}\BibitemShut {NoStop}%
\bibitem [{\citenamefont
  {Karolewski}(2001)}]{Karolewski-RadiatEffDefectS-2001}%
  \BibitemOpen
  \bibfield  {author} {\bibinfo {author} {\bibfnamefont {M.~A.}\ \bibnamefont
  {Karolewski}},\ }\href {\doibase 10.1080/10420150108211842} {\bibfield
  {journal} {\bibinfo  {journal} {Radiat. Eff. Defect S.}\ }\textbf {\bibinfo
  {volume} {153}},\ \bibinfo {pages} {239} (\bibinfo {year}
  {2001})}\BibitemShut {NoStop}%
\bibitem [{\citenamefont {Johnson}(1989)}]{Johnson-PRB-1989}%
  \BibitemOpen
  \bibfield  {author} {\bibinfo {author} {\bibfnamefont {R.~A.}\ \bibnamefont
  {Johnson}},\ }\href {\doibase 10.1103/PhysRevB.39.12554} {\bibfield
  {journal} {\bibinfo  {journal} {Phys. Rev. B}\ }\textbf {\bibinfo {volume}
  {39}},\ \bibinfo {pages} {12554} (\bibinfo {year} {1989})}\BibitemShut
  {NoStop}%
\bibitem [{\citenamefont {Politano}\ \emph {et~al.}(2013)\citenamefont
  {Politano}, \citenamefont {Chiarello}, \citenamefont {Benedek}, \citenamefont
  {Chulkov},\ and\ \citenamefont {Echenique}}]{Politano-SurfSciRep-2013}%
  \BibitemOpen
  \bibfield  {author} {\bibinfo {author} {\bibfnamefont {A.}~\bibnamefont
  {Politano}}, \bibinfo {author} {\bibfnamefont {G.}~\bibnamefont {Chiarello}},
  \bibinfo {author} {\bibfnamefont {G.}~\bibnamefont {Benedek}}, \bibinfo
  {author} {\bibfnamefont {E.}~\bibnamefont {Chulkov}}, \ and\ \bibinfo
  {author} {\bibfnamefont {P.}~\bibnamefont {Echenique}},\ }\href {\doibase
  10.1016/j.surfrep.2013.07.001} {\bibfield  {journal} {\bibinfo  {journal}
  {Surf. Sci. Rep.}\ }\textbf {\bibinfo {volume} {68}},\ \bibinfo {pages} {305}
  (\bibinfo {year} {2013})}\BibitemShut {NoStop}%
\bibitem [{\citenamefont {Rusina}\ \emph {et~al.}(2005)\citenamefont {Rusina},
  \citenamefont {Eremeev}, \citenamefont {Borisova}, \citenamefont
  {Sklyadneva},\ and\ \citenamefont {Chulkov}}]{Rusina-PRB-2005}%
  \BibitemOpen
  \bibfield  {author} {\bibinfo {author} {\bibfnamefont {G.~G.}\ \bibnamefont
  {Rusina}}, \bibinfo {author} {\bibfnamefont {S.~V.}\ \bibnamefont {Eremeev}},
  \bibinfo {author} {\bibfnamefont {S.~D.}\ \bibnamefont {Borisova}}, \bibinfo
  {author} {\bibfnamefont {I.~Y.}\ \bibnamefont {Sklyadneva}}, \ and\ \bibinfo
  {author} {\bibfnamefont {E.~V.}\ \bibnamefont {Chulkov}},\ }\href {\doibase
  10.1103/PhysRevB.71.245401} {\bibfield  {journal} {\bibinfo  {journal} {Phys.
  Rev. B}\ }\textbf {\bibinfo {volume} {71}},\ \bibinfo {pages} {245401}
  (\bibinfo {year} {2005})}\BibitemShut {NoStop}%
\bibitem [{\citenamefont {Rusina}\ \emph {et~al.}(2007)\citenamefont {Rusina},
  \citenamefont {Eremeev}, \citenamefont {Borisova}, \citenamefont
  {Sklyadneva}, \citenamefont {Echenique},\ and\ \citenamefont
  {Chulkov}}]{Rusina-JPhysCondensMatter-2007}%
  \BibitemOpen
  \bibfield  {author} {\bibinfo {author} {\bibfnamefont {G.~G.}\ \bibnamefont
  {Rusina}}, \bibinfo {author} {\bibfnamefont {S.~V.}\ \bibnamefont {Eremeev}},
  \bibinfo {author} {\bibfnamefont {S.~D.}\ \bibnamefont {Borisova}}, \bibinfo
  {author} {\bibfnamefont {I.~Y.}\ \bibnamefont {Sklyadneva}}, \bibinfo
  {author} {\bibfnamefont {P.~M.}\ \bibnamefont {Echenique}}, \ and\ \bibinfo
  {author} {\bibfnamefont {E.~V.}\ \bibnamefont {Chulkov}},\ }\href {\doibase
  10.1088/0953-8984/19/26/266005} {\bibfield  {journal} {\bibinfo  {journal}
  {J. Phys.: Condens. Matter}\ }\textbf {\bibinfo {volume} {19}},\ \bibinfo
  {pages} {266005} (\bibinfo {year} {2007})}\BibitemShut {NoStop}%
\bibitem [{\citenamefont {Rusina}\ \emph {et~al.}(2008)\citenamefont {Rusina},
  \citenamefont {Eremeev}, \citenamefont {Echenique}, \citenamefont {Benedek},
  \citenamefont {Borisova},\ and\ \citenamefont
  {Chulkov}}]{Rusina-JPhysCondensMatter-2008}%
  \BibitemOpen
  \bibfield  {author} {\bibinfo {author} {\bibfnamefont {G.~G.}\ \bibnamefont
  {Rusina}}, \bibinfo {author} {\bibfnamefont {S.~V.}\ \bibnamefont {Eremeev}},
  \bibinfo {author} {\bibfnamefont {P.~M.}\ \bibnamefont {Echenique}}, \bibinfo
  {author} {\bibfnamefont {G.}~\bibnamefont {Benedek}}, \bibinfo {author}
  {\bibfnamefont {S.~D.}\ \bibnamefont {Borisova}}, \ and\ \bibinfo {author}
  {\bibfnamefont {E.~V.}\ \bibnamefont {Chulkov}},\ }\href {\doibase
  10.1088/0953-8984/20/22/224007} {\bibfield  {journal} {\bibinfo  {journal}
  {J. Phys.: Condens. Matter}\ }\textbf {\bibinfo {volume} {20}},\ \bibinfo
  {pages} {224007} (\bibinfo {year} {2008})}\BibitemShut {NoStop}%
\bibitem [{\citenamefont {Borisova}\ \emph {et~al.}(2006)\citenamefont
  {Borisova}, \citenamefont {Rusina}, \citenamefont {Eremeev}, \citenamefont
  {Benedek}, \citenamefont {Echenique}, \citenamefont {Sklyadneva},\ and\
  \citenamefont {Chulkov}}]{Borisova-PRB-2006}%
  \BibitemOpen
  \bibfield  {author} {\bibinfo {author} {\bibfnamefont {S.~D.}\ \bibnamefont
  {Borisova}}, \bibinfo {author} {\bibfnamefont {G.~G.}\ \bibnamefont
  {Rusina}}, \bibinfo {author} {\bibfnamefont {S.~V.}\ \bibnamefont {Eremeev}},
  \bibinfo {author} {\bibfnamefont {G.}~\bibnamefont {Benedek}}, \bibinfo
  {author} {\bibfnamefont {P.~M.}\ \bibnamefont {Echenique}}, \bibinfo {author}
  {\bibfnamefont {I.~Y.}\ \bibnamefont {Sklyadneva}}, \ and\ \bibinfo {author}
  {\bibfnamefont {E.~V.}\ \bibnamefont {Chulkov}},\ }\href {\doibase
  10.1103/PhysRevB.74.165412} {\bibfield  {journal} {\bibinfo  {journal} {Phys.
  Rev. B}\ }\textbf {\bibinfo {volume} {74}},\ \bibinfo {pages} {1} (\bibinfo
  {year} {2006})}\BibitemShut {NoStop}%
\bibitem [{\citenamefont {Rusina}\ \emph {et~al.}(2012)\citenamefont {Rusina},
  \citenamefont {Eremeev}, \citenamefont {Borisova},\ and\ \citenamefont
  {Chulkov}}]{Rusina-JPhysCondensMatter-2012}%
  \BibitemOpen
  \bibfield  {author} {\bibinfo {author} {\bibfnamefont {G.~G.}\ \bibnamefont
  {Rusina}}, \bibinfo {author} {\bibfnamefont {S.~V.}\ \bibnamefont {Eremeev}},
  \bibinfo {author} {\bibfnamefont {S.~D.}\ \bibnamefont {Borisova}}, \ and\
  \bibinfo {author} {\bibfnamefont {E.~V.}\ \bibnamefont {Chulkov}},\ }\href
  {\doibase 10.1088/0953-8984/24/10/104003} {\bibfield  {journal} {\bibinfo
  {journal} {J. Phys.: Condens. Matter}\ }\textbf {\bibinfo {volume} {24}},\
  \bibinfo {pages} {104003} (\bibinfo {year} {2012})}\BibitemShut {NoStop}%
\bibitem [{\citenamefont {Perdew}\ \emph {et~al.}(1996)\citenamefont {Perdew},
  \citenamefont {Burke},\ and\ \citenamefont {Ernzerhof}}]{Perdew-PRL-1996}%
  \BibitemOpen
  \bibfield  {author} {\bibinfo {author} {\bibfnamefont {J.~P.}\ \bibnamefont
  {Perdew}}, \bibinfo {author} {\bibfnamefont {K.}~\bibnamefont {Burke}}, \
  and\ \bibinfo {author} {\bibfnamefont {M.}~\bibnamefont {Ernzerhof}},\ }\href
  {\doibase 10.1103/PhysRevLett.77.3865} {\bibfield  {journal} {\bibinfo
  {journal} {Phys. Rev. Lett.}\ }\textbf {\bibinfo {volume} {77}},\ \bibinfo
  {pages} {3865} (\bibinfo {year} {1996})}\BibitemShut {NoStop}%
\bibitem [{\citenamefont {Perdew}\ \emph {et~al.}(1997)\citenamefont {Perdew},
  \citenamefont {Burke},\ and\ \citenamefont {Ernzerhof}}]{Perdew-PRL-1997}%
  \BibitemOpen
  \bibfield  {author} {\bibinfo {author} {\bibfnamefont {J.~P.}\ \bibnamefont
  {Perdew}}, \bibinfo {author} {\bibfnamefont {K.}~\bibnamefont {Burke}}, \
  and\ \bibinfo {author} {\bibfnamefont {M.}~\bibnamefont {Ernzerhof}},\ }\href
  {\doibase 10.1103/PhysRevLett.78.1396} {\bibfield  {journal} {\bibinfo
  {journal} {Phys. Rev. Lett.}\ }\textbf {\bibinfo {volume} {78}},\ \bibinfo
  {pages} {1396} (\bibinfo {year} {1997})}\BibitemShut {NoStop}%
\bibitem [{\citenamefont {Kresse}\ and\ \citenamefont
  {Hafner}(1993)}]{kresse1993}%
  \BibitemOpen
  \bibfield  {author} {\bibinfo {author} {\bibfnamefont {G.}~\bibnamefont
  {Kresse}}\ and\ \bibinfo {author} {\bibfnamefont {J.}~\bibnamefont
  {Hafner}},\ }\href {\doibase 10.1103/PhysRevB.47.558} {\bibfield  {journal}
  {\bibinfo  {journal} {Phys. Rev. B}\ }\textbf {\bibinfo {volume} {47}},\
  \bibinfo {pages} {558} (\bibinfo {year} {1993})}\BibitemShut {NoStop}%
\bibitem [{\citenamefont {Kresse}\ and\ \citenamefont
  {Furthm{\"{u}}ller}(1996{\natexlab{a}})}]{kresse1996a}%
  \BibitemOpen
  \bibfield  {author} {\bibinfo {author} {\bibfnamefont {G.}~\bibnamefont
  {Kresse}}\ and\ \bibinfo {author} {\bibfnamefont {J.}~\bibnamefont
  {Furthm{\"{u}}ller}},\ }\href {\doibase 10.1016/0927-0256(96)00008-0}
  {\bibfield  {journal} {\bibinfo  {journal} {Comput. Mat. Sci.}\ }\textbf
  {\bibinfo {volume} {6}},\ \bibinfo {pages} {15} (\bibinfo {year}
  {1996}{\natexlab{a}})}\BibitemShut {NoStop}%
\bibitem [{\citenamefont {Kresse}\ and\ \citenamefont
  {Furthm{\"{u}}ller}(1996{\natexlab{b}})}]{kresse1996b}%
  \BibitemOpen
  \bibfield  {author} {\bibinfo {author} {\bibfnamefont {G.}~\bibnamefont
  {Kresse}}\ and\ \bibinfo {author} {\bibfnamefont {J.}~\bibnamefont
  {Furthm{\"{u}}ller}},\ }\href {\doibase 10.1103/PhysRevB.54.11169} {\bibfield
   {journal} {\bibinfo  {journal} {Phys. Rev. B}\ }\textbf {\bibinfo {volume}
  {54}},\ \bibinfo {pages} {11169} (\bibinfo {year}
  {1996}{\natexlab{b}})}\BibitemShut {NoStop}%
\bibitem [{\citenamefont {Kresse}\ and\ \citenamefont
  {Joubert}(1999)}]{kresse1999}%
  \BibitemOpen
  \bibfield  {author} {\bibinfo {author} {\bibfnamefont {G.}~\bibnamefont
  {Kresse}}\ and\ \bibinfo {author} {\bibfnamefont {D.}~\bibnamefont
  {Joubert}},\ }\href {\doibase 10.1103/PhysRevB.59.1758} {\bibfield  {journal}
  {\bibinfo  {journal} {Phys. Rev. B}\ }\textbf {\bibinfo {volume} {59}},\
  \bibinfo {pages} {1758} (\bibinfo {year} {1999})}\BibitemShut {NoStop}%
\bibitem [{\citenamefont {Parlinski}\ \emph {et~al.}(1997)\citenamefont
  {Parlinski}, \citenamefont {Li},\ and\ \citenamefont
  {Kawazoe}}]{Parlinski-PRL-1997}%
  \BibitemOpen
  \bibfield  {author} {\bibinfo {author} {\bibfnamefont {K.}~\bibnamefont
  {Parlinski}}, \bibinfo {author} {\bibfnamefont {Z.~Q.}\ \bibnamefont {Li}}, \
  and\ \bibinfo {author} {\bibfnamefont {Y.}~\bibnamefont {Kawazoe}},\ }\href
  {\doibase 10.1103/PhysRevLett.78.4063} {\bibfield  {journal} {\bibinfo
  {journal} {Phys. Rev. Lett.}\ }\textbf {\bibinfo {volume} {78}},\ \bibinfo
  {pages} {4063} (\bibinfo {year} {1997})}\BibitemShut {NoStop}%
\bibitem [{\citenamefont {Togo}\ and\ \citenamefont
  {Tanaka}(2015)}]{Togo-ScriptaMaterialia-2015}%
  \BibitemOpen
  \bibfield  {author} {\bibinfo {author} {\bibfnamefont {A.}~\bibnamefont
  {Togo}}\ and\ \bibinfo {author} {\bibfnamefont {I.}~\bibnamefont {Tanaka}},\
  }\href {\doibase 10.1016/j.scriptamat.2015.07.021} {\bibfield  {journal}
  {\bibinfo  {journal} {Scr. Mater.}\ }\textbf {\bibinfo {volume} {108}},\
  \bibinfo {pages} {1} (\bibinfo {year} {2015})}\BibitemShut {NoStop}%
\bibitem [{\citenamefont {Benedek}\ \emph {et~al.}(1993)\citenamefont
  {Benedek}, \citenamefont {Ellis}, \citenamefont {Luo}, \citenamefont
  {Reichmuth}, \citenamefont {Ruggerone},\ and\ \citenamefont
  {Toennies}}]{Benedek-PRB-1993}%
  \BibitemOpen
  \bibfield  {author} {\bibinfo {author} {\bibfnamefont {G.}~\bibnamefont
  {Benedek}}, \bibinfo {author} {\bibfnamefont {J.}~\bibnamefont {Ellis}},
  \bibinfo {author} {\bibfnamefont {N.~S.}\ \bibnamefont {Luo}}, \bibinfo
  {author} {\bibfnamefont {A.}~\bibnamefont {Reichmuth}}, \bibinfo {author}
  {\bibfnamefont {P.}~\bibnamefont {Ruggerone}}, \ and\ \bibinfo {author}
  {\bibfnamefont {J.~P.}\ \bibnamefont {Toennies}},\ }\href {\doibase
  10.1103/PhysRevB.48.4917} {\bibfield  {journal} {\bibinfo  {journal} {Phys.
  Rev. B}\ }\textbf {\bibinfo {volume} {48}},\ \bibinfo {pages} {4917}
  (\bibinfo {year} {1993})}\BibitemShut {NoStop}%
\bibitem [{\citenamefont {Wuttig}\ \emph
  {et~al.}(1986{\natexlab{a}})\citenamefont {Wuttig}, \citenamefont {Franchy},\
  and\ \citenamefont {Ibach}}]{Wuttig-ZPhysB-1986}%
  \BibitemOpen
  \bibfield  {author} {\bibinfo {author} {\bibfnamefont {M.}~\bibnamefont
  {Wuttig}}, \bibinfo {author} {\bibfnamefont {R.}~\bibnamefont {Franchy}}, \
  and\ \bibinfo {author} {\bibfnamefont {H.}~\bibnamefont {Ibach}},\ }\href
  {\doibase 10.1007/BF01308401} {\bibfield  {journal} {\bibinfo  {journal} {Z.
  Phys. B Con. Mat.}\ }\textbf {\bibinfo {volume} {65}},\ \bibinfo {pages} {71}
  (\bibinfo {year} {1986}{\natexlab{a}})}\BibitemShut {NoStop}%
\bibitem [{\citenamefont {Wuttig}\ \emph
  {et~al.}(1986{\natexlab{b}})\citenamefont {Wuttig}, \citenamefont {Franchy},\
  and\ \citenamefont {Ibach}}]{Wuttig-SolidStateCommun-1986}%
  \BibitemOpen
  \bibfield  {author} {\bibinfo {author} {\bibfnamefont {M.}~\bibnamefont
  {Wuttig}}, \bibinfo {author} {\bibfnamefont {R.}~\bibnamefont {Franchy}}, \
  and\ \bibinfo {author} {\bibfnamefont {H.}~\bibnamefont {Ibach}},\ }\href
  {\doibase 10.1016/0038-1098(86)90488-6} {\bibfield  {journal} {\bibinfo
  {journal} {Solid State Commun.}\ }\textbf {\bibinfo {volume} {57}},\ \bibinfo
  {pages} {445} (\bibinfo {year} {1986}{\natexlab{b}})}\BibitemShut {NoStop}%
\bibitem [{\citenamefont {Nelson}\ \emph {et~al.}(1988)\citenamefont {Nelson},
  \citenamefont {Sowa},\ and\ \citenamefont {Daw}}]{Nelson-PRL-1988}%
  \BibitemOpen
  \bibfield  {author} {\bibinfo {author} {\bibfnamefont {J.~S.}\ \bibnamefont
  {Nelson}}, \bibinfo {author} {\bibfnamefont {E.~C.}\ \bibnamefont {Sowa}}, \
  and\ \bibinfo {author} {\bibfnamefont {M.~S.}\ \bibnamefont {Daw}},\ }\href
  {\doibase 10.1103/PhysRevLett.61.1977} {\bibfield  {journal} {\bibinfo
  {journal} {Phys. Rev. Lett.}\ }\textbf {\bibinfo {volume} {61}},\ \bibinfo
  {pages} {1977} (\bibinfo {year} {1988})}\BibitemShut {NoStop}%
\bibitem [{\citenamefont {Fratesi}(2009)}]{Fratesi-PRB-2009}%
  \BibitemOpen
  \bibfield  {author} {\bibinfo {author} {\bibfnamefont {G.}~\bibnamefont
  {Fratesi}},\ }\href {\doibase 10.1103/PhysRevB.80.045422} {\bibfield
  {journal} {\bibinfo  {journal} {Phys. Rev. B}\ }\textbf {\bibinfo {volume}
  {80}},\ \bibinfo {pages} {045422} (\bibinfo {year} {2009})}\BibitemShut
  {NoStop}%
\bibitem [{\citenamefont {Senet}\ \emph {et~al.}(1999)\citenamefont {Senet},
  \citenamefont {Toennies},\ and\ \citenamefont
  {Witte}}]{Senet-ChemPhysLett-1999}%
  \BibitemOpen
  \bibfield  {author} {\bibinfo {author} {\bibfnamefont {P.}~\bibnamefont
  {Senet}}, \bibinfo {author} {\bibfnamefont {J.}~\bibnamefont {Toennies}}, \
  and\ \bibinfo {author} {\bibfnamefont {G.}~\bibnamefont {Witte}},\ }\href
  {\doibase 10.1016/S0009-2614(98)01314-1} {\bibfield  {journal} {\bibinfo
  {journal} {Chem. Phys. Lett.}\ }\textbf {\bibinfo {volume} {299}},\ \bibinfo
  {pages} {389} (\bibinfo {year} {1999})}\BibitemShut {NoStop}%
\bibitem [{\citenamefont {Graham}(2003)}]{Graham-SurfSciRep-2003}%
  \BibitemOpen
  \bibfield  {author} {\bibinfo {author} {\bibfnamefont {A.~P.}\ \bibnamefont
  {Graham}},\ }\href {\doibase 10.1016/S0167-5729(03)00012-8} {\bibfield
  {journal} {\bibinfo  {journal} {Surf. Sci. Rep.}\ }\textbf {\bibinfo {volume}
  {49}},\ \bibinfo {pages} {115} (\bibinfo {year} {2003})}\BibitemShut
  {NoStop}%
\bibitem [{\citenamefont {Astaldi}\ \emph {et~al.}(1990)\citenamefont
  {Astaldi}, \citenamefont {Rudolf},\ and\ \citenamefont
  {Modesti}}]{Astaldi-SolidStateCommun-1990}%
  \BibitemOpen
  \bibfield  {author} {\bibinfo {author} {\bibfnamefont {C.}~\bibnamefont
  {Astaldi}}, \bibinfo {author} {\bibfnamefont {P.}~\bibnamefont {Rudolf}}, \
  and\ \bibinfo {author} {\bibfnamefont {S.}~\bibnamefont {Modesti}},\ }\href
  {\doibase 10.1016/0038-1098(90)90396-S} {\bibfield  {journal} {\bibinfo
  {journal} {Solid State Commun.}\ }\textbf {\bibinfo {volume} {75}},\ \bibinfo
  {pages} {847} (\bibinfo {year} {1990})}\BibitemShut {NoStop}%
\bibitem [{\citenamefont {Graham}\ \emph
  {et~al.}(1997{\natexlab{a}})\citenamefont {Graham}, \citenamefont {Hofmann},
  \citenamefont {Toennies}, \citenamefont {Chen},\ and\ \citenamefont
  {Ying}}]{Graham-PRL-1997}%
  \BibitemOpen
  \bibfield  {author} {\bibinfo {author} {\bibfnamefont {A.~P.}\ \bibnamefont
  {Graham}}, \bibinfo {author} {\bibfnamefont {F.}~\bibnamefont {Hofmann}},
  \bibinfo {author} {\bibfnamefont {J.~P.}\ \bibnamefont {Toennies}}, \bibinfo
  {author} {\bibfnamefont {L.~Y.}\ \bibnamefont {Chen}}, \ and\ \bibinfo
  {author} {\bibfnamefont {S.~C.}\ \bibnamefont {Ying}},\ }\href {\doibase
  10.1103/PhysRevLett.78.3900} {\bibfield  {journal} {\bibinfo  {journal}
  {Phys. Rev. Lett.}\ }\textbf {\bibinfo {volume} {78}},\ \bibinfo {pages}
  {3900} (\bibinfo {year} {1997}{\natexlab{a}})}\BibitemShut {NoStop}%
\bibitem [{\citenamefont {Graham}\ \emph
  {et~al.}(1997{\natexlab{b}})\citenamefont {Graham}, \citenamefont {Hofmann},
  \citenamefont {Toennies}, \citenamefont {Chen},\ and\ \citenamefont
  {Ying}}]{Graham-PRB-1997-1}%
  \BibitemOpen
  \bibfield  {author} {\bibinfo {author} {\bibfnamefont {A.~P.}\ \bibnamefont
  {Graham}}, \bibinfo {author} {\bibfnamefont {F.}~\bibnamefont {Hofmann}},
  \bibinfo {author} {\bibfnamefont {J.~P.}\ \bibnamefont {Toennies}}, \bibinfo
  {author} {\bibfnamefont {L.~Y.}\ \bibnamefont {Chen}}, \ and\ \bibinfo
  {author} {\bibfnamefont {S.~C.}\ \bibnamefont {Ying}},\ }\href {\doibase
  10.1103/PhysRevB.56.10567} {\bibfield  {journal} {\bibinfo  {journal} {Phys.
  Rev. B}\ }\textbf {\bibinfo {volume} {56}},\ \bibinfo {pages} {10567}
  (\bibinfo {year} {1997}{\natexlab{b}})}\BibitemShut {NoStop}%
\bibitem [{\citenamefont {Persson}\ and\ \citenamefont
  {Ryberg}(1985)}]{Persson-PRB-1985}%
  \BibitemOpen
  \bibfield  {author} {\bibinfo {author} {\bibfnamefont {B.~N.~J.}\
  \bibnamefont {Persson}}\ and\ \bibinfo {author} {\bibfnamefont
  {R.}~\bibnamefont {Ryberg}},\ }\href {\doibase 10.1103/PhysRevB.32.3586}
  {\bibfield  {journal} {\bibinfo  {journal} {Phys. Rev. B}\ }\textbf {\bibinfo
  {volume} {32}},\ \bibinfo {pages} {3586} (\bibinfo {year}
  {1985})}\BibitemShut {NoStop}%
\bibitem [{\citenamefont {Lewis}\ \emph {et~al.}(1998)\citenamefont {Lewis},
  \citenamefont {Pykhtin}, \citenamefont {Mele},\ and\ \citenamefont
  {Rappe}}]{Lewis-JCP-1998}%
  \BibitemOpen
  \bibfield  {author} {\bibinfo {author} {\bibfnamefont {S.~P.}\ \bibnamefont
  {Lewis}}, \bibinfo {author} {\bibfnamefont {M.~V.}\ \bibnamefont {Pykhtin}},
  \bibinfo {author} {\bibfnamefont {E.~J.}\ \bibnamefont {Mele}}, \ and\
  \bibinfo {author} {\bibfnamefont {A.~M.}\ \bibnamefont {Rappe}},\ }\href
  {\doibase 10.1063/1.475478} {\bibfield  {journal} {\bibinfo  {journal} {J.
  Chem. Phys.}\ }\textbf {\bibinfo {volume} {108}},\ \bibinfo {pages} {1157}
  (\bibinfo {year} {1998})}\BibitemShut {NoStop}%
\bibitem [{\citenamefont {Pykhtin}\ \emph {et~al.}(1998)\citenamefont
  {Pykhtin}, \citenamefont {Lewis}, \citenamefont {Mele},\ and\ \citenamefont
  {Rappe}}]{Pykhtin-PRL-1998}%
  \BibitemOpen
  \bibfield  {author} {\bibinfo {author} {\bibfnamefont {M.~V.}\ \bibnamefont
  {Pykhtin}}, \bibinfo {author} {\bibfnamefont {S.~P.}\ \bibnamefont {Lewis}},
  \bibinfo {author} {\bibfnamefont {E.~J.}\ \bibnamefont {Mele}}, \ and\
  \bibinfo {author} {\bibfnamefont {A.~M.}\ \bibnamefont {Rappe}},\ }\href
  {\doibase 10.1103/PhysRevLett.81.5940} {\bibfield  {journal} {\bibinfo
  {journal} {Phys. Rev. Lett.}\ }\textbf {\bibinfo {volume} {81}},\ \bibinfo
  {pages} {5940} (\bibinfo {year} {1998})}\BibitemShut {NoStop}%
\bibitem [{\citenamefont {Persson}\ \emph {et~al.}(1999)\citenamefont
  {Persson}, \citenamefont {Tosatti}, \citenamefont {Fuhrmann}, \citenamefont
  {Witte},\ and\ \citenamefont {W{\"{o}}ll}}]{Persson-PRB-1999}%
  \BibitemOpen
  \bibfield  {author} {\bibinfo {author} {\bibfnamefont {B.~N.~J.}\
  \bibnamefont {Persson}}, \bibinfo {author} {\bibfnamefont {E.}~\bibnamefont
  {Tosatti}}, \bibinfo {author} {\bibfnamefont {D.}~\bibnamefont {Fuhrmann}},
  \bibinfo {author} {\bibfnamefont {G.}~\bibnamefont {Witte}}, \ and\ \bibinfo
  {author} {\bibfnamefont {C.}~\bibnamefont {W{\"{o}}ll}},\ }\href {\doibase
  10.1103/PhysRevB.59.11777} {\bibfield  {journal} {\bibinfo  {journal} {Phys.
  Rev. B}\ }\textbf {\bibinfo {volume} {59}},\ \bibinfo {pages} {11777}
  (\bibinfo {year} {1999})}\BibitemShut {NoStop}%
\bibitem [{\citenamefont {West}\ and\ \citenamefont
  {Estreicher}(2006)}]{West-PRL-2006}%
  \BibitemOpen
  \bibfield  {author} {\bibinfo {author} {\bibfnamefont {D.}~\bibnamefont
  {West}}\ and\ \bibinfo {author} {\bibfnamefont {S.~K.}\ \bibnamefont
  {Estreicher}},\ }\href {\doibase 10.1103/PhysRevLett.96.115504} {\bibfield
  {journal} {\bibinfo  {journal} {Phys. Rev. Lett.}\ }\textbf {\bibinfo
  {volume} {96}},\ \bibinfo {pages} {115504} (\bibinfo {year}
  {2006})}\BibitemShut {NoStop}%
\bibitem [{\citenamefont {West}\ and\ \citenamefont
  {Estreicher}(2007)}]{West-PRB-2007}%
  \BibitemOpen
  \bibfield  {author} {\bibinfo {author} {\bibfnamefont {D.}~\bibnamefont
  {West}}\ and\ \bibinfo {author} {\bibfnamefont {S.~K.}\ \bibnamefont
  {Estreicher}},\ }\href {\doibase 10.1103/PhysRevB.75.075206} {\bibfield
  {journal} {\bibinfo  {journal} {Phys. Rev. B}\ }\textbf {\bibinfo {volume}
  {75}},\ \bibinfo {pages} {075206} (\bibinfo {year} {2007})}\BibitemShut
  {NoStop}%
\bibitem [{\citenamefont {Gibbons}\ and\ \citenamefont
  {Estreicher}(2009)}]{Gibbons-PRL-2009}%
  \BibitemOpen
  \bibfield  {author} {\bibinfo {author} {\bibfnamefont {T.~M.}\ \bibnamefont
  {Gibbons}}\ and\ \bibinfo {author} {\bibfnamefont {S.~K.}\ \bibnamefont
  {Estreicher}},\ }\href {\doibase 10.1103/PhysRevLett.102.255502} {\bibfield
  {journal} {\bibinfo  {journal} {Phys. Rev. Lett.}\ }\textbf {\bibinfo
  {volume} {102}},\ \bibinfo {pages} {26} (\bibinfo {year} {2009})}\BibitemShut
  {NoStop}%
\bibitem [{\citenamefont {Estreicher}\ \emph {et~al.}(2009)\citenamefont
  {Estreicher}, \citenamefont {Backlund}, \citenamefont {Gibbons},\ and\
  \citenamefont {Docaj}}]{Estreicher-ModellingSimulMaterSciEng-2009}%
  \BibitemOpen
  \bibfield  {author} {\bibinfo {author} {\bibfnamefont {S.~K.}\ \bibnamefont
  {Estreicher}}, \bibinfo {author} {\bibfnamefont {D.}~\bibnamefont
  {Backlund}}, \bibinfo {author} {\bibfnamefont {T.~M.}\ \bibnamefont
  {Gibbons}}, \ and\ \bibinfo {author} {\bibfnamefont {A.}~\bibnamefont
  {Docaj}},\ }\href {\doibase 10.1088/0965-0393/17/8/084006} {\bibfield
  {journal} {\bibinfo  {journal} {Modelling Simul. Mater. Sci. Eng.}\ }\textbf
  {\bibinfo {volume} {17}},\ \bibinfo {pages} {084006} (\bibinfo {year}
  {2009})}\BibitemShut {NoStop}%
\bibitem [{\citenamefont {McGaughey}\ and\ \citenamefont
  {Kaviany}(2004)}]{McGaughey-PRB-2004}%
  \BibitemOpen
  \bibfield  {author} {\bibinfo {author} {\bibfnamefont {A.~J.~H.}\
  \bibnamefont {McGaughey}}\ and\ \bibinfo {author} {\bibfnamefont
  {M.}~\bibnamefont {Kaviany}},\ }\href {\doibase 10.1103/PhysRevB.69.094303}
  {\bibfield  {journal} {\bibinfo  {journal} {Phys. Rev. B}\ }\textbf {\bibinfo
  {volume} {69}},\ \bibinfo {pages} {1} (\bibinfo {year} {2004})}\BibitemShut
  {NoStop}%
\bibitem [{\citenamefont {Meyer}(2011)}]{Meyer-PhDThesis}%
  \BibitemOpen
  \bibfield  {author} {\bibinfo {author} {\bibfnamefont {J.}~\bibnamefont
  {Meyer}},\ }\emph {\bibinfo {title} {Ab initio Modeling of Energy Dissipation
  during Chemical Reactions at Transition Metal Surfaces}},\ \href@noop {}
  {Ph.D. thesis},\ \bibinfo  {school} {Freie Universit\"at Berlin} (\bibinfo
  {year} {2011})\BibitemShut {NoStop}%
\bibitem [{\citenamefont {Bukas}\ and\ \citenamefont
  {Reuter}(2016)}]{Bukas-PRL-2016}%
  \BibitemOpen
  \bibfield  {author} {\bibinfo {author} {\bibfnamefont {V.~J.}\ \bibnamefont
  {Bukas}}\ and\ \bibinfo {author} {\bibfnamefont {K.}~\bibnamefont {Reuter}},\
  }\href {\doibase 10.1103/PhysRevLett.117.146101} {\bibfield  {journal}
  {\bibinfo  {journal} {Phys. Rev. Lett.}\ }\textbf {\bibinfo {volume} {117}},\
  \bibinfo {pages} {146101} (\bibinfo {year} {2016})}\BibitemShut {NoStop}%
\bibitem [{\citenamefont {Bukas}\ and\ \citenamefont
  {Reuter}(2017)}]{Bukas-JCP-2017}%
  \BibitemOpen
  \bibfield  {author} {\bibinfo {author} {\bibfnamefont {V.~J.}\ \bibnamefont
  {Bukas}}\ and\ \bibinfo {author} {\bibfnamefont {K.}~\bibnamefont {Reuter}},\
  }\href {\doibase 10.1063/1.4973244} {\bibfield  {journal} {\bibinfo
  {journal} {J. Chem. Phys.}\ }\textbf {\bibinfo {volume} {146}},\ \bibinfo
  {pages} {14702} (\bibinfo {year} {2017})}\BibitemShut {NoStop}%
\bibitem [{\citenamefont {Tully}(1980)}]{Tully-JCP-1980}%
  \BibitemOpen
  \bibfield  {author} {\bibinfo {author} {\bibfnamefont {J.~C.}\ \bibnamefont
  {Tully}},\ }\href {\doibase 10.1063/1.440287} {\bibfield  {journal} {\bibinfo
   {journal} {J. Chem. Phys.}\ }\textbf {\bibinfo {volume} {73}},\ \bibinfo
  {pages} {1975} (\bibinfo {year} {1980})}\BibitemShut {NoStop}%
\bibitem [{\citenamefont {Tully}\ \emph {et~al.}(1993)\citenamefont {Tully},
  \citenamefont {Gomez},\ and\ \citenamefont
  {Head-Gordon}}]{Tully-JVacSciTechnol-1993}%
  \BibitemOpen
  \bibfield  {author} {\bibinfo {author} {\bibfnamefont {J.~C.}\ \bibnamefont
  {Tully}}, \bibinfo {author} {\bibfnamefont {M.}~\bibnamefont {Gomez}}, \ and\
  \bibinfo {author} {\bibfnamefont {M.}~\bibnamefont {Head-Gordon}},\ }\href
  {\doibase 10.1116/1.578522} {\bibfield  {journal} {\bibinfo  {journal} {J.
  Vac. Sci. Technol. A}\ }\textbf {\bibinfo {volume} {11}},\ \bibinfo {pages}
  {1914} (\bibinfo {year} {1993})}\BibitemShut {NoStop}%
\bibitem [{\citenamefont {{Head-Gordon}}\ and\ \citenamefont
  {Tully}(1995)}]{HeadGordon-JCP-1995}%
  \BibitemOpen
  \bibfield  {author} {\bibinfo {author} {\bibfnamefont {M.}~\bibnamefont
  {{Head-Gordon}}}\ and\ \bibinfo {author} {\bibfnamefont {J.~C.}\ \bibnamefont
  {Tully}},\ }\href {\doibase 10.1063/1.469915} {\bibfield  {journal} {\bibinfo
   {journal} {J. Chem. Phys.}\ }\textbf {\bibinfo {volume} {103}},\ \bibinfo
  {pages} {10137} (\bibinfo {year} {1995})}\BibitemShut {NoStop}%
\bibitem [{\citenamefont {Juaristi}\ \emph {et~al.}(2008)\citenamefont
  {Juaristi}, \citenamefont {Alducin}, \citenamefont {D{\'i}ez~Mui{\~n}o},
  \citenamefont {Busnengo},\ and\ \citenamefont {Salin}}]{Juaristi-PRL-2008}%
  \BibitemOpen
  \bibfield  {author} {\bibinfo {author} {\bibfnamefont {J.~I.}\ \bibnamefont
  {Juaristi}}, \bibinfo {author} {\bibfnamefont {M.}~\bibnamefont {Alducin}},
  \bibinfo {author} {\bibfnamefont {R.}~\bibnamefont {D{\'i}ez~Mui{\~n}o}},
  \bibinfo {author} {\bibfnamefont {H.~F.}\ \bibnamefont {Busnengo}}, \ and\
  \bibinfo {author} {\bibfnamefont {A.}~\bibnamefont {Salin}},\ }\href
  {\doibase 10.1103/PhysRevLett.100.116102} {\bibfield  {journal} {\bibinfo
  {journal} {Phys. Rev. Lett.}\ }\textbf {\bibinfo {volume} {100}},\ \bibinfo
  {pages} {116102} (\bibinfo {year} {2008})}\BibitemShut {NoStop}%
\bibitem [{\citenamefont {Rittmeyer}\ \emph {et~al.}(2016)\citenamefont
  {Rittmeyer}, \citenamefont {Ward}, \citenamefont {G{\"{u}}tlein},
  \citenamefont {Ellis}, \citenamefont {Allison},\ and\ \citenamefont
  {Reuter}}]{Rittmeyer-PRL-2016}%
  \BibitemOpen
  \bibfield  {author} {\bibinfo {author} {\bibfnamefont {S.~P.}\ \bibnamefont
  {Rittmeyer}}, \bibinfo {author} {\bibfnamefont {D.~J.}\ \bibnamefont {Ward}},
  \bibinfo {author} {\bibfnamefont {P.}~\bibnamefont {G{\"{u}}tlein}}, \bibinfo
  {author} {\bibfnamefont {J.}~\bibnamefont {Ellis}}, \bibinfo {author}
  {\bibfnamefont {W.}~\bibnamefont {Allison}}, \ and\ \bibinfo {author}
  {\bibfnamefont {K.}~\bibnamefont {Reuter}},\ }\href {\doibase
  10.1103/PhysRevLett.117.196001} {\bibfield  {journal} {\bibinfo  {journal}
  {Phys. Rev. Lett.}\ }\textbf {\bibinfo {volume} {117}},\ \bibinfo {pages}
  {196001} (\bibinfo {year} {2016})}\BibitemShut {NoStop}%
\bibitem [{\citenamefont {Rittmeyer}\ \emph {et~al.}(2015)\citenamefont
  {Rittmeyer}, \citenamefont {Meyer}, \citenamefont {Juaristi},\ and\
  \citenamefont {Reuter}}]{Rittmeyer-PRL-2015}%
  \BibitemOpen
  \bibfield  {author} {\bibinfo {author} {\bibfnamefont {S.~P.}\ \bibnamefont
  {Rittmeyer}}, \bibinfo {author} {\bibfnamefont {J.}~\bibnamefont {Meyer}},
  \bibinfo {author} {\bibfnamefont {J.~I.}\ \bibnamefont {Juaristi}}, \ and\
  \bibinfo {author} {\bibfnamefont {K.}~\bibnamefont {Reuter}},\ }\href
  {\doibase 10.1103/PhysRevLett.115.046102} {\bibfield  {journal} {\bibinfo
  {journal} {Phys. Rev. Lett.}\ }\textbf {\bibinfo {volume} {115}},\ \bibinfo
  {pages} {046102} (\bibinfo {year} {2015})}\BibitemShut {NoStop}%
\end{thebibliography}%
\end{document}


%
%
\title{%
\texorpdfstring{\textit{Supplemental Material:}\\ Temperature-Dependent Lifetimes of Low-Frequency Adsorbate Modes from Non-Equilibrium Molecular Dynamics Simulations}{Supplemental Material: Temperature-Dependent Lifetimes of Low-Frequency Adsorbate Modes from Non-Equilibrium Molecular Dynamics Simulations}
}
\author{Francesco Nattino}
\affiliation{Leiden Institute of Chemistry, Leiden University, Gorlaeus Laboratories, P.O. Box 9502, 2300 RA Leiden, The Netherlands}
\author{J\"org Meyer}
\affiliation{Leiden Institute of Chemistry, Leiden University, Gorlaeus Laboratories, P.O. Box 9502, 2300 RA Leiden, The Netherlands}
\date{\today} 
\maketitle
\tableofcontents
\pagebreak
%
%
\section{The Potential \label{sec:potential}}

\subsection{Parametetrization \label{sub:parametrization}}

According to the embedded atom method (EAM) \cite{Daw-PRL-1983, Daw-PRB-1984}, 
the potential energy $U$ of a system including $N$ atoms is given by:
\begin{equation}
U = \sum_{i}^N\left(F_{\alpha\left(i\right)}\left(\rho_i\right) + \frac{1}{2}\sum_{0<r_{ij}<r_c}^{N_c}\phi_{\alpha\left(i\right)\beta\left(j\right)}(r_{ij})\right). \label{eq:pot_energy}
\end{equation}
Greek characters ($\alpha, \beta, ...$) are used here to represent atom types  
while Latin characters ($i, j, ...$) identify atom indices. $r_{ij}$ is the distance between 
the atom $i$, of type $\alpha(i)$, and the atom $j$, of type $\beta(j)$. 
The electron density in which the atom $i$ is embedded, $\rho_i$, is given by:
\begin{equation}
\rho_i = \sum_{0<r_{ij}<r_c}^{N_c} f_{\alpha\left(j\right)}\left(r_{ij}\right). \label{eq:embedding_density}
\end{equation}
The sums in Equations \ref{eq:pot_energy} and \ref{eq:embedding_density}
normally extend to the $N_c$ atoms within a given cutoff radius $r_c$ from the
atom $i$, with $r_c$ typically corresponding to few nearest neighbor distances.    
Starting from the original work of Daw and Baskes \cite{Daw-PRL-1983, Daw-PRB-1984}, 
various functional forms have been proposed for the density function $f_{\alpha}$, 
the embedding function $F_{\alpha}$ and the pair potential $\phi_{\alpha\beta}$.  
In the current work, the functional forms of $f_{\alpha}$ 
and $F_{\alpha}$, as well as $\phi_{\alpha\beta}$ for the $\alpha=\beta$ case, 
have been obtained by straightforwardly recasting  
the expressions proposed by Cleri and Rosato \cite{Cleri-PRB-1993} in a tight-binding formalism
in an EAM potential form:
\begin{subequations}
\begin{align}
f_{\alpha}(r_{ij})&=\xi^2e^{-2q(r_{ij}/r_0 - 1)}\label{eq:TB-f}\\
F_{\alpha}(\rho)&=-\sqrt{\rho}\label{eq:TB-F}\\
\phi_{\alpha\beta}(r_{ij})&=2Ae^{-p(r_{ij}/r_0 - 1)}\label{eq:TB-phi}
\end{align}
\end{subequations}
For the density and embedding functions as well as for the Cu-Cu and the Na-Na pair potentials, 
we have employed the parametrization provided by Karolewski \cite{Karolewski-RadiatEffDefectS-2001},
who has fitted the five parameters in Equations \ref{eq:TB-f}-\ref{eq:TB-phi}
(i.e. $\xi,q,r_0,p$ and $A$) to experimental data for a wide range of pure metals. In particular, the fitted potentials 
exactly reproduce the experimental lattice constants and cohesive energies, while they have been optimized 
in order to best match experimental elastic constants and vacancy formation energies. The lattice constant values 
chosen by Karolewski in the fit of the EAM parameters for Cu and Na are 3.6147 \AA\ and 4.2906 \AA, 
respectively \cite{Karolewski-RadiatEffDefectS-2001}. The EAM parameters employed 
in the current work are reported in Table \ref{tab:EAMparameters}. The value  
employed for the cutoff radius $r_c = 4.021$ \AA\ corresponds to a distance intermediate between the second and 
the third nearest neighbor distances for bulk Cu.

\begin{table}
\begin{ruledtabular}
\begin{tabular}{ccdd}
Parameter	&	Unit & \text{Cu (fcc)} & \text{Na (bcc)} \tabularnewline
\hline 
$\xi$ &	eV	&	1.2355	&	0.4083 \tabularnewline
$q$		&	-		&	2.3197	&	1.7477 \tabularnewline
$r_0$	&	\AA	&	2.5560	&	3.7158 \tabularnewline
$p$		&	-		&	11.1832	&	7.8536 \tabularnewline
$A$		&	eV	&	0.0783	&	0.0353 \tabularnewline
\hline 
$r_c$	&	\AA	&	4.0209	&	6.4014 \tabularnewline
\end{tabular}
\end{ruledtabular}
\caption{EAM parameters and cutoff radius 
employed for Cu and Na in the current work, as originally proposed for a tight-binding formalism by
Karolewski \cite{Karolewski-RadiatEffDefectS-2001}.}
\label{tab:EAMparameters}
\end{table}

The Cu-Na two-body potential $\phi_{\mathrm{Cu Na}}$ has been constructed from the 
pair potentials for Cu-Cu and Na-Na interactions using the alloy model 
proposed by Johnson \cite{Johnson-PRB-1989}:
\begin{equation}
\phi_{\mathrm{Cu Na}}(r) = \frac{1}{2} \left[ \frac{f_{\mathrm{Na}}(r)}{f_{\mathrm{Cu}}(r)} \phi_{\mathrm{Cu Cu}}(r) + 
      \frac{f_{\mathrm{Cu}}(r)}{f_{\mathrm{Na}}(r)} \phi_{\mathrm{Na Na}}(r)  \right]. \label{eq:Johnson-pair-potential}
\end{equation}
This model, which has been shown to accurately predict the heat of solution 
for binary alloys of Cu, Ag, Au, Ni and Pt \cite{Johnson-PRB-1989}, 
has also been applied to the study of numerous alkali metal 
adatom-transition metal surface systems \cite{Politano-SurfSciRep-2013}: 
Li, Na, and K on Al(111) \cite{Rusina-PRB-2005}, Li and Na on Al(100) \cite{Rusina-JPhysCondensMatter-2007}, Li and Na 
on Cu(100) \cite{Rusina-JPhysCondensMatter-2008}, Na on Cu(111) \cite{Borisova-PRB-2006}
and K on Pt(111) \cite{Rusina-JPhysCondensMatter-2012}. 
Unfortunately, the exact EAM parametrizations employed in these references 
are not accessible, thus demanding the present potential construction and 
validation efforts.
Overall, the use of EAM potentials in combination with the Johnson alloy model 
described above has represented a semi-empirical and computationally affordable method 
to calculate accurate phonon dispersions and other vibrational 
properties for systems whose unit cell size would have made 
higher level calculations (e.g. density functional perturbation theory, DFPT) 
extremely challenging. 

\subsection{Validation \label{sub:validation}}

\subsubsection{Cu bulk and Cu(100) \label{ss:1}}

We have thoroughly investigated the quality of the EAM potential employed 
by comparing computed observables to
available experimental data or higher level calculations. 
For what concerns the substrate vibrational properties, the parameters employed here
provide a very good description of the phonon dispersion for both Cu bulk and Cu(100). 
EAM phonon dispersions and density of states (DOSs) have been calculated and 
compared to corresponding density functional theory (DFT) simulations and experimental data. 
DFT calculations have been based on the PBE exchange-correlation functional \cite{Perdew-PRL-1996, Perdew-PRL-1997}
as implemented in the VASP code \cite{kresse1993, kresse1996a, kresse1996b, kresse1999}.  
Computational parameters include a 500 eV energy cutoff for the plane wave expansion, PAW potentials \cite{kresse1999} 
to represent core electrons, a Fermi smearing with width parameter $\sigma=0.1$ eV to facilitate convergence and 
a k-point sampling corresponding to a 18x18x18 $\Gamma$-centered grid in the bulk primitive cell. 
The equilibrium lattice constant $a=3.636$ \AA\ has been obtained with the described setup.

\begin{figure}
\begin{centering}
	\includegraphics[width=0.7\columnwidth]{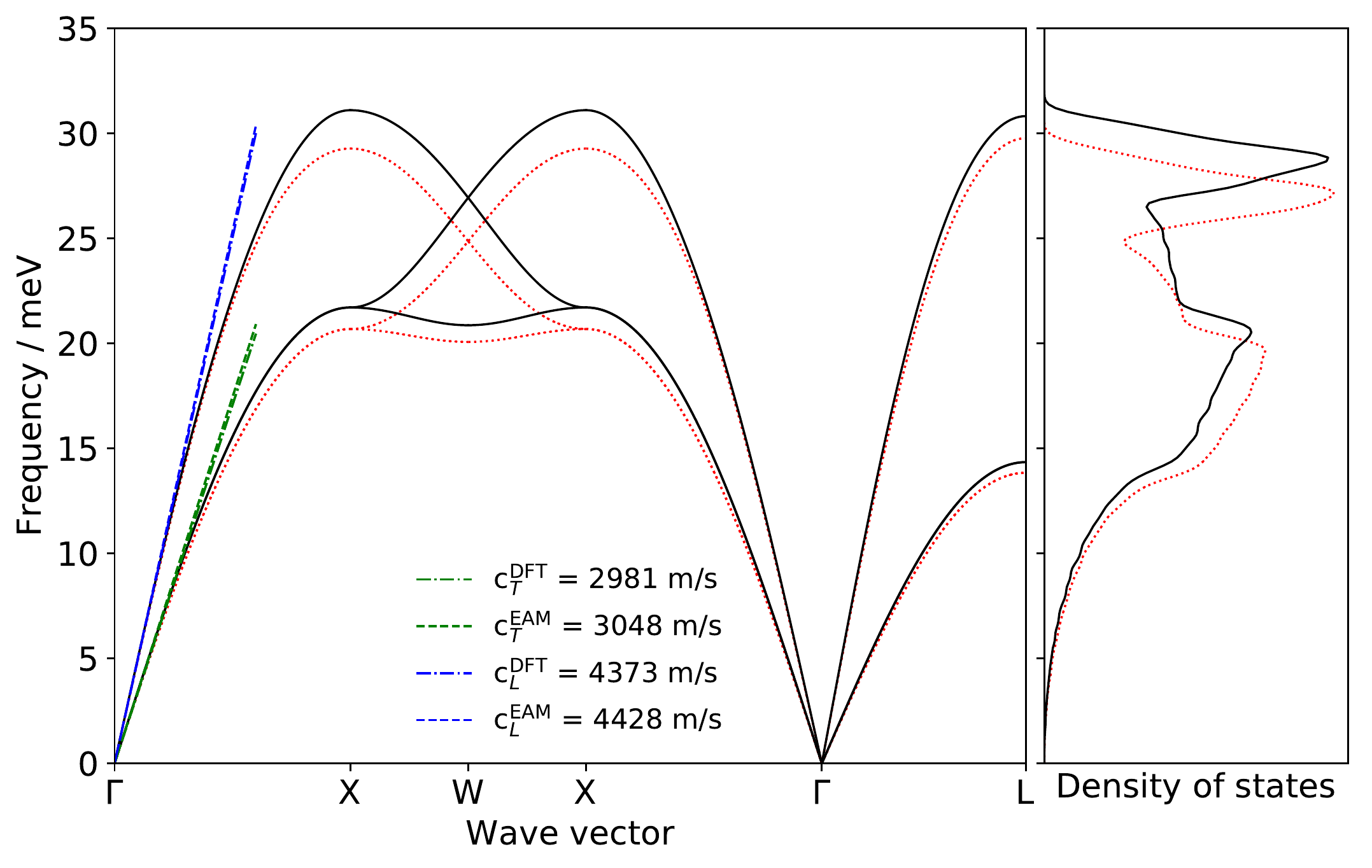}\\
\end{centering}
\caption{Phonon dispersion (left) and DOSs (right) for fcc bulk copper, calculated with DFT (dotted red lines) 
and with the EAM potential (solid black lines). 
Linear dispersions that correspond to group velocities calculated for the three acoustic branches close to the $\Gamma$ point 
are also shown as dashed blue (green) lines for the EAM potential (DFT) calculations.}
\label{fig:bulk_dispersion}
\end{figure}

All phonon calculations have been performed with a finite difference approach \cite{Parlinski-PRL-1997}
as implemented in the PHONOPY package \cite{Togo-ScriptaMaterialia-2015}, 
using a displacement size corresponding to 0.01 \AA.
For what concerns bulk calculations, the force constant matrix has been calculated 
in a 3x3x3 multiple of the conventional fcc unit cell for the DFT calculations 
after checking convergence of the phonon dispersion and the density of states 
with respect to the supercell size. 
Considering the significantly lower computational cost required for the evaluation of 
EAM energies (and forces), a larger 6x6x6 multiple of 
the bulk primitive cell has been employed in
the EAM phonon calculations. Figure \ref{fig:bulk_dispersion} shows the comparison between 
DFT and EAM bulk phonon dispersions and DOSs. The two methods agree 
well in both band shapes and maximum phonon frequency, the difference being smaller than 2 meV. 
Good agreement has been found for 
the group velocities of the transversal and the longitudinal acoustic 
phonon branch, as calculated along the [100] crystallographic direction at a \textbf{q}-point 
very close to $\Gamma$ (see Figure \ref{fig:bulk_dispersion}).

\begin{figure}
\begin{centering}
	\includegraphics[width=0.5\columnwidth]{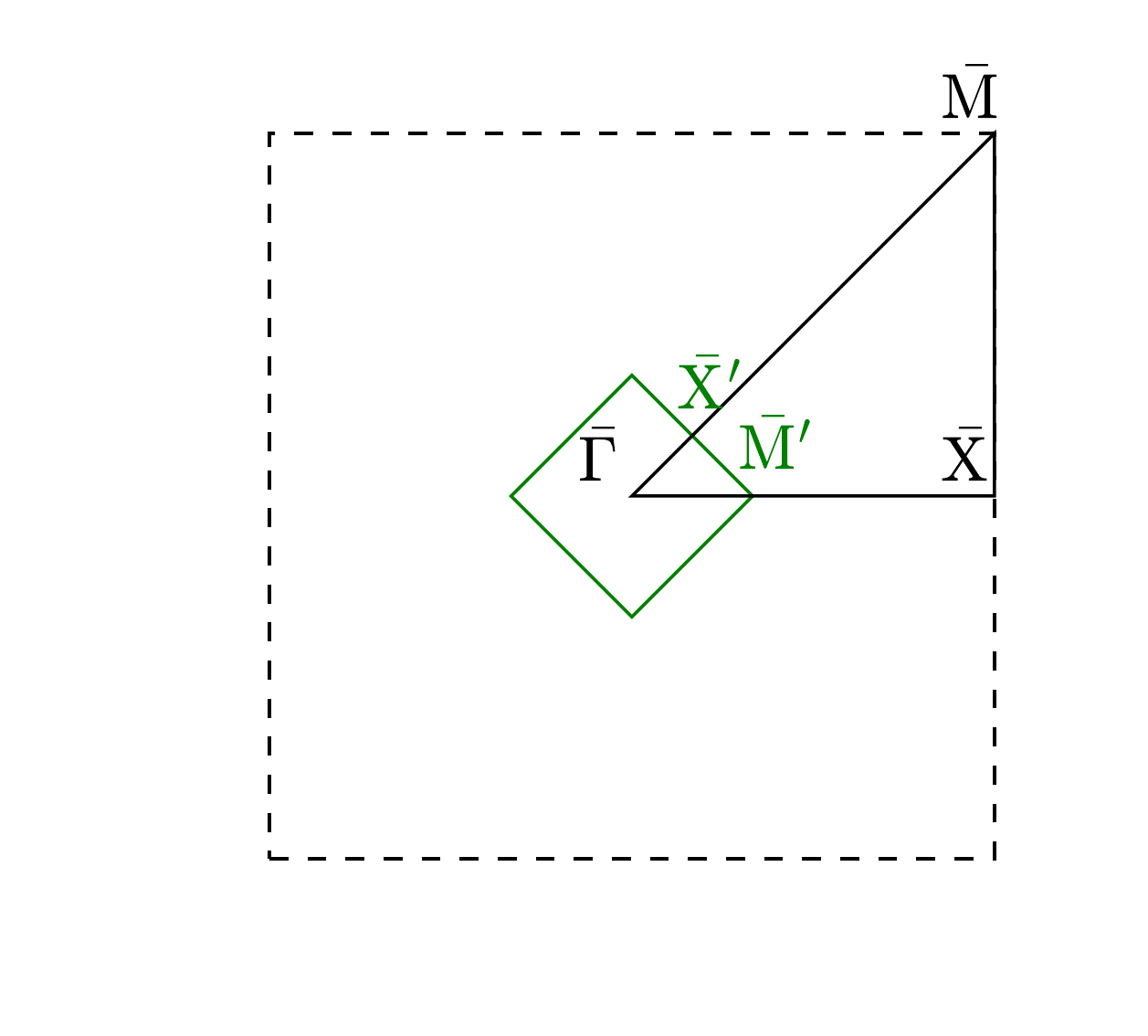}\\
\end{centering}
\caption{Surface Brillouin zone and corresponding high symmetry points for a clean Cu(100) surface (black) 
and for the Cu(100)-Na-c$(6\times6)$ overlayer structure (green).}
\label{fig:brillouin_zone}
\end{figure}

\begin{figure}
\begin{centering}
	\includegraphics[width=0.7\columnwidth]{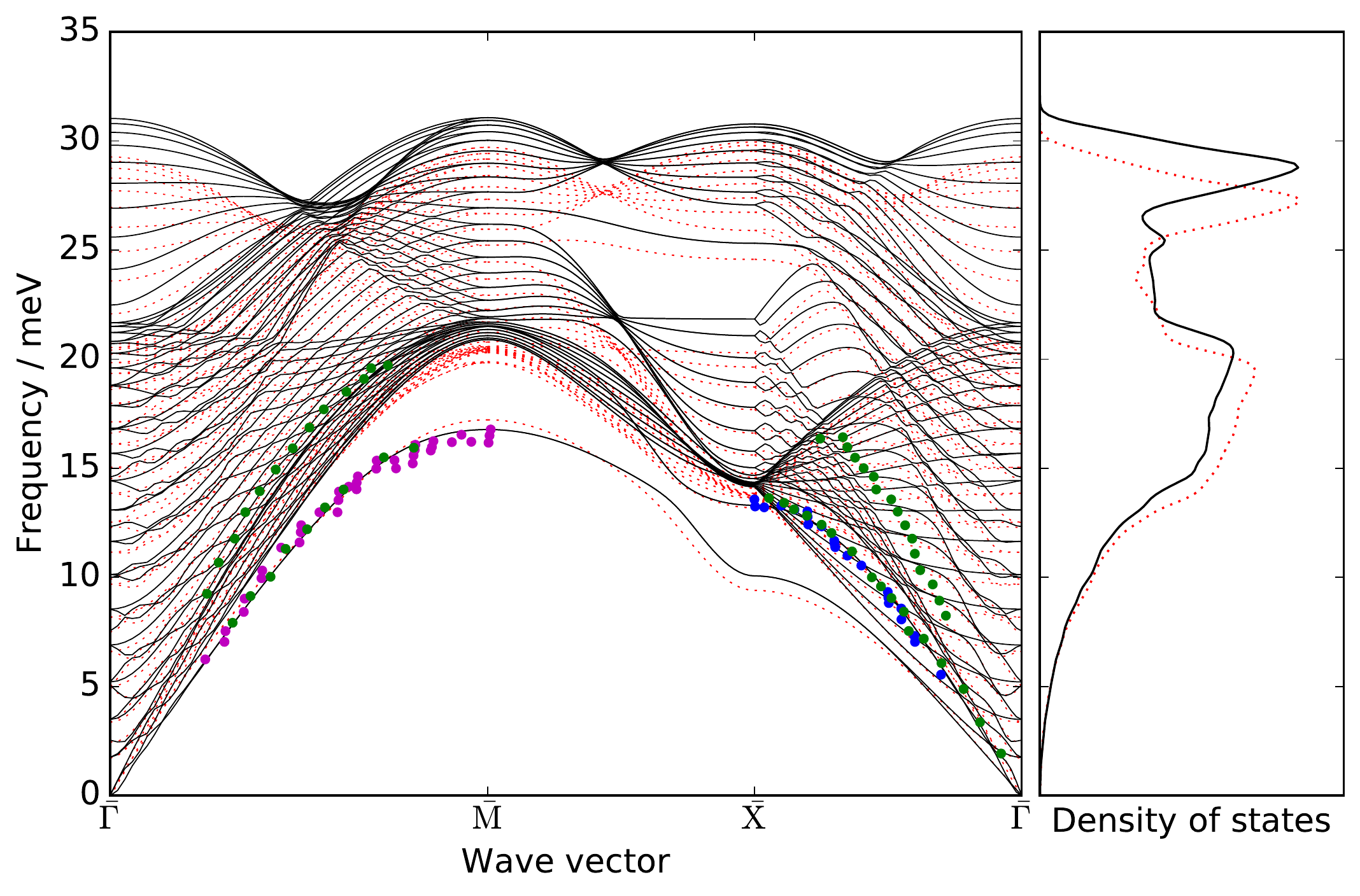}\\
\end{centering}
\caption{Phonon dispersion (left) and DOS (right) for a clean Cu(100) slab, calculated with DFT (dotted red lines) 
and with the EAM potential (solid black lines). Available experimental data have also been included, from HAS experiments (green symbols, 
from Ref. \cite{Benedek-PRB-1993}) and from EELS measurements (magenta and blue symbols, 
from Ref. \cite{Wuttig-ZPhysB-1986} and \cite{Wuttig-SolidStateCommun-1986}, respectively).}
\label{fig:Cu100_dispersion}
\end{figure}

Surface phonon dispersions have been calculated for both EAM and DFT  
using a relaxed slab with 20 atomic layers.  A  6x6x1 (3x3x1) multiple of the primitive cell
has been employed for the EAM (DFT) calculation of the force constant matrix.  
The Cu(100) surface Brillouin zone is schematically
depicted in Figure  \ref{fig:brillouin_zone} for convenience. As for bulk calculations,
EAM surface phonon dispersions agree well with DFT data, with 
a very modest upward shift (less than 2 meV, see Figure \ref{fig:Cu100_dispersion}).
Good agreement is also found between EAM calculations and experimental data, 
as determined through helium atom scattering (HAS) experiments 
\cite{Benedek-PRB-1993} and electron energy loss spectroscopy (EELS)
\cite{Wuttig-ZPhysB-1986, Wuttig-SolidStateCommun-1986}. Note that in the 
$\overline{\mathrm{X}\Gamma}$ segment the lowest frequency surface phonon mode 
was not experimentally observed because of the phonon polarization 
vector being perpendicular to the sagittal plane that characterizes the 
scattering geometry employed, as already noted in Ref. \cite{Nelson-PRL-1988}.

\subsubsection{Na overlayers on Cu(100) \label{ss:2}}

The constructed EAM potential yields excellent 
agreement with experimental and theoretical data for the Na+Cu adsorption system, as described in this section. 
First of all, the potential predicts the four-fold hollow site as stable adsorption site for 
Na adatoms on Cu(100), in agreement with experimental observations and DFT studies \cite{Fratesi-PRB-2009}. 
For what concerns the vibrational properties of the coupled adsorbate-substrate system, 
calculated phonon dispersions have been reported by Chulkov and coworkers 
for a different Cu surface, i.e. Cu(111) \cite{Borisova-PRB-2006, Politano-SurfSciRep-2013}. 
We have considered the Na on Cu(111) adsorption structures from 
Refs. \cite{Borisova-PRB-2006, Politano-SurfSciRep-2013} 
that are in the same coverage range as the structures in the current work (i.e. up to 25\% ML). 
By using the constructed EAM potential, we have been able to reproduce essentially 
all the features of the reported phonon dispersions, like 
the frequency of the Na frustrated translation (FT) mode and its lack of dispersion.

\begin{figure}
\begin{centering}
	\includegraphics[width=0.7\columnwidth]{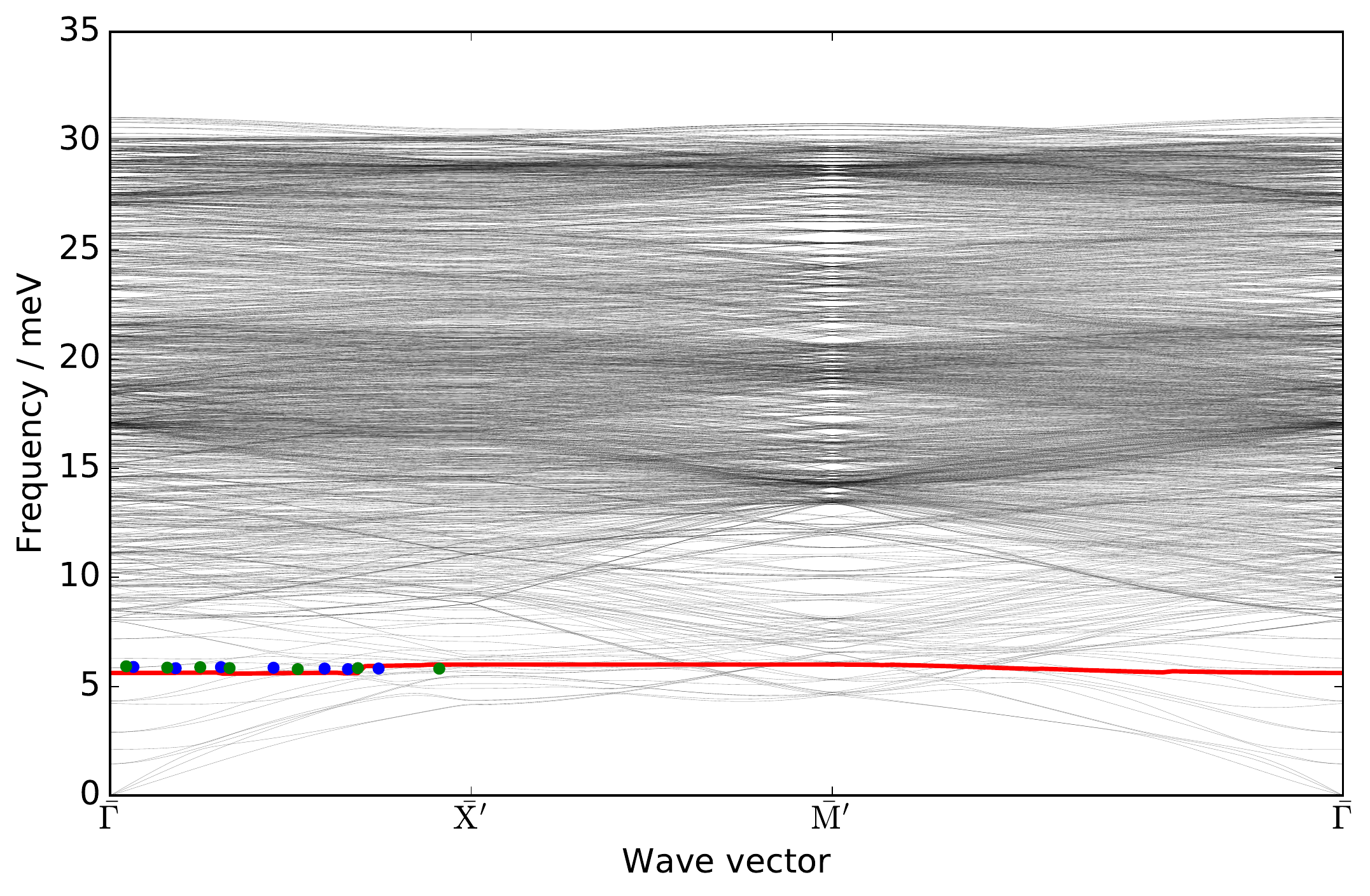}\\
\end{centering}
\caption{Phonon dispersion for the Cu(100)-Na-c$(6\times6)$ overlayer structure calculated with the EAM potential. 
The dispersion of the phonon mode with largest Na FT-mode character is shown in red. 
Available HAS experimental data for the dispersion of Na FT-mode have also been included  
(blue and green symbols, from Ref. \cite{Graham-SurfSciRep-2003} and \cite{Senet-ChemPhysLett-1999}, respectively).}
\label{fig:NaCu100_dispersion}
\end{figure}

In Figure \ref{fig:NaCu100_dispersion} we show the phonon dispersion calculated 
for the Na-c$(6\times6)$ overlayer structure, corresponding to a coverage $\theta=1/18\approx5.6\%$. 
Figure 1C in the main paper shows a (real space) representation of the overlayer structure, while the 
corresponding irreducible surface Brillouin zone is sketched in Figure \ref{fig:brillouin_zone}.
A supercell corresponding to a 6x6x1 multiple of the substrate unit cell 
has been employed to calculate the force constant matrix of the adsorption system
for all the overlayer structures considered in the current work.
In Figure \ref{fig:NaCu100_dispersion} we have highlighted the phonon band corresponding 
to the mode with highest Na FT mode character and we compare it 
to data from HAS experiments that have investigated the dispersion of the 
Na FT-mode along the [100] crystallographic direction 
\cite{Senet-ChemPhysLett-1999, Graham-SurfSciRep-2003}. 
Note that the precise Na overlayer structure in these experiments is not known, but the reported coverage is 
close to the one employed in our simulations (i.e. 5.6\%): 5\% in Ref. \cite{Senet-ChemPhysLett-1999} 
and 4.8\% in Ref. \cite{Graham-SurfSciRep-2003}.
The constructed potential reproduces very well the frequency of this phonon mode 
over the entire SBZ. We note that due to the cutoff distance $r_c$ employed
all the overlayer structures considered here yield dispersion-free Na FT-modes, which is in 
agreement with experimental data \cite{Senet-ChemPhysLett-1999, Graham-SurfSciRep-2003}.   
Our EAM potential predicts a value for the frequency of the surface-Na stretching mode 
in the range 16-18 meV, also in good agreement with the experimentally determined 
value of 18 meV \cite{Astaldi-SolidStateCommun-1990}.
 
\begin{figure}
\begin{centering}
	\includegraphics[width=0.7\columnwidth]{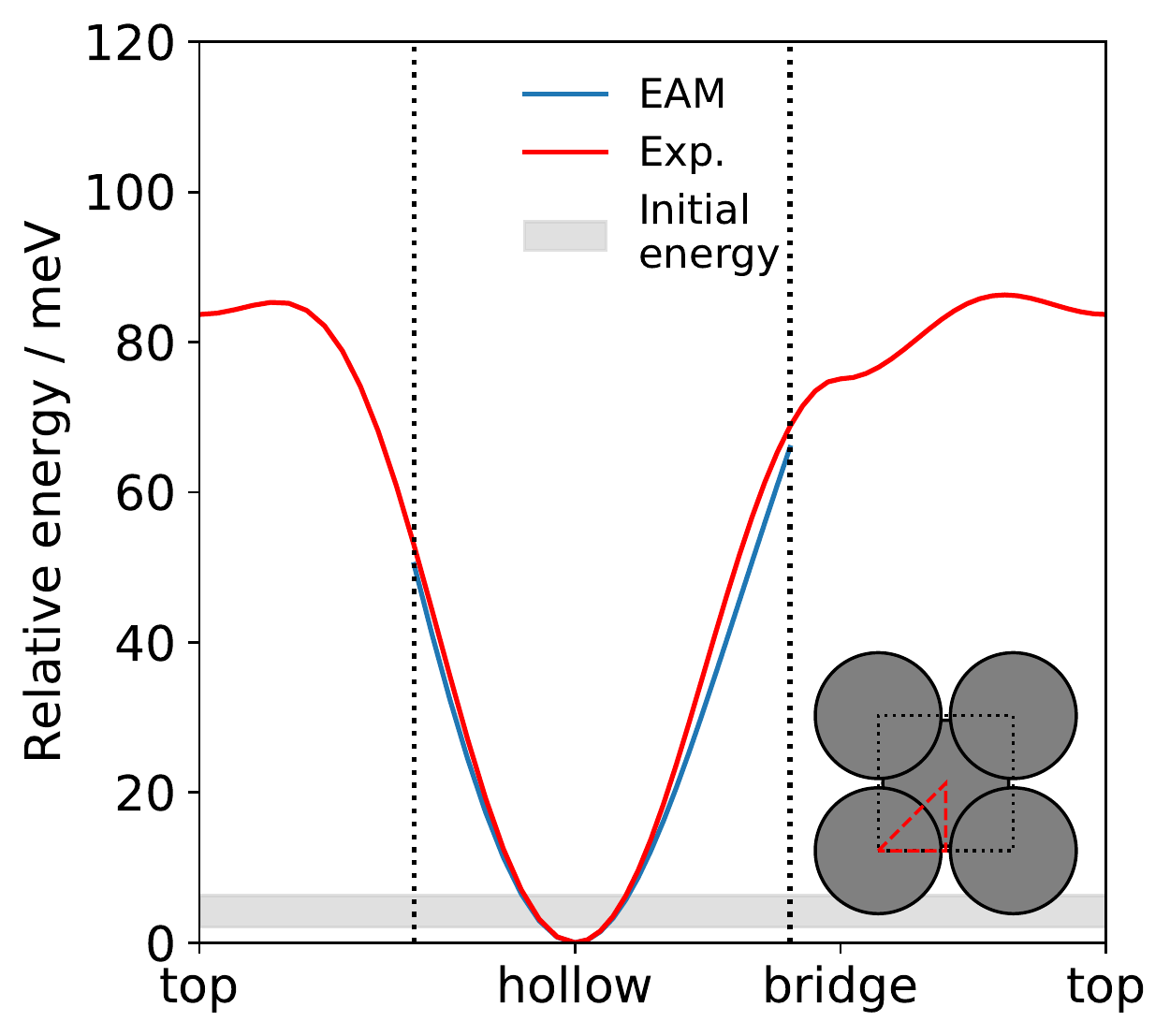}\\
\end{centering}
\caption{A 1D cut of the experiment-based 2D PES describing the lateral motion of 
Na on Cu(100)\cite{Graham-PRL-1997, Graham-PRB-1997-1} is shown in red. 
The EAM energy calculated along the same path after optimizing 
the $Z$ coordinate of the Na atom while freezing all the other degrees of freedom is plotted in blue.   
Vertical lines indicate the continuity interval of the potential around the four-fold hollow site.
The gray bar illustrates the excitation energy range per adsorbate atom that is considered in the 
current work.
The path considered is shown as a red dashed line in the inset, where we have also sketched 
primitive cell of the Cu(100) surface (dotted black line).}
\label{fig:PES}
\end{figure}

In Figure \ref{fig:PES} we finally compare the interaction potential between Na and Cu(100) 
as calculated with the EAM potential and as predicted through a two dimensional (2D) 
potential energy surface (PES) constructed on the basis of HAS 
experimental data \cite{Graham-PRL-1997, Graham-PRB-1997-1}. 
In the EAM calculations  the $Z$ coordinate of the Na atom has been relaxed while the lateral coordinates 
have been kept fixed along the path investigated. The Cu(100) surface has been kept frozen 
in its equilibrium configuration. Equal PESs are obtained in the primitive cell of all 
overlayer structures due to the lack of Na-Na lateral interactions as a consequence of the 
cutoff distance $r_c$ employed. The agreement between the EAM potential and the 
experimentally-obtained PES in the relevant region around the most stable adsorption site 
(i.e. the four fold hollow site) is excellent. Note that the EAM potential does not yield 
to a continuous representation of the Na-Cu(100) interaction potential over the whole adsorbate
coordinate space. In fact, discontinuities arise for large atomic displacements from the equilibrium configuration, 
following the change of the number of Cu atoms in the neighbor shell of a Na atom if the same 
moves from the four-fold-coordinated hollow site to e.g. the two-fold-coordinated bridge site. 
This is a consequence of the rather short cutoff radius with which EAM
potentials are typically fitted \cite{Karolewski-RadiatEffDefectS-2001} (see also Equation \ref{eq:pot_energy} 
and \ref{eq:embedding_density}). 
In Figure \ref{fig:PES} the PES calculated with the EAM potential is only plotted within the `continuity interval',
omitting the jumps occurring at the interval boundaries. 
This limitation, however, does not constitute 
an issue for the study of adsorbate mode lifetimes at low temperatures, as carried out here. 
In fact, for this purpose the only relevant portion of the phase space is the area in the vicinity 
of the equilibrium adsorption configuration, which is extremely well described here.   
In addition, trajectories in which a Na atom hops to a neighboring hollow site due to thermal diffusion are
discarded when studying the de-excitation of the Na FT-mode, 
regardless of the discontinuities in the potential (see also the following section).   

\section{Effective Continuum Model \label{sec:ECM}}

In order to estimate the vibrational lifetime of low-frequency adsorbate modes on metal surfaces, 
Persson and Rydberg originally proposed an analytical model based on an elastic-continuum description of 
the substrate \cite{Persson-PRB-1985}. 
The model has been subsequently extended by Rappe and coworkers \cite{Lewis-JCP-1998, Pykhtin-PRL-1998}. 
The input for such elastic continuum model (ECM) consists of
a very limited set of parameters, presented in Table \ref{tab:ECM}
together with the values employed here.
Note that the values of all these parameters have been 
estimated on the basis of the EAM potential that we have constructed, 
as also employed in the non-equilibrium molecular dynamics (NEMD) calculations.
The lifetimes calculated with the ECM and the ones
extracted from the NEMD calculations are therefore consistently based on the 
same system parameterization, allowing for a rigorous comparison of the two approaches. 

\begin{table}
\begin{ruledtabular}
\begin{tabular}{llr}
Parameter &	& Value \tabularnewline
\hline 
Adsorbate frequency 										& $\omega$	& 6.01~\text{meV} \tabularnewline
Adsorbate mass													& $m$ 			& 22.99~\text{amu} \tabularnewline
Bulk lattice constant										& $a$				& 3.615~\text{\AA} \tabularnewline
Substrate density												& $\rho$		& 8.937~\text{g/cm$^3$} \tabularnewline
Substrate transversal speed of sound		& $c_{T}$		& 3048~\text{m/s} \tabularnewline
Substrate longitudinal speed of sound 	&	$c_{L}$		& 4428~\text{m/s} \tabularnewline
\end{tabular}
\end{ruledtabular}
\caption{EAM-based parameters employed in the ECM lifetime calculations. The adlayer unit cell area
$A$ can be straightforwardly calculated for each overlayer structure from the bulk lattice constant
$a$, provided in the table. The values of $c_{T,L}$ have been calculated as the group velocities of the 
acoustic modes close to $\Gamma$ (see also Figure \ref{fig:bulk_dispersion}).}
\label{tab:ECM}
\end{table}

In particular, according to the ECM,
the lifetime of a low-frequency adsorbate mode can be calculated according to the following 
expression \cite{Lewis-JCP-1998, Pykhtin-PRL-1998}:
\begin{equation}
\tau_{\mathrm{ECM}} = \frac{(1+R(\omega))^2 + I^2(\omega)}{\omega I(\omega)},
\end{equation}
where $\omega$ is the frequency of the adsorbate mode of interest, and 
$R(\omega)$ and $I(\omega)$ are the real and the imaginary part, respectively, of the following discretized integral:
\begin{equation}
T(\omega)  = \frac{m\omega^2\theta}{A} \sum_{\textbf{q}_i}D(c_{T,L},\rho;\textbf{q}_i,\omega).
\end{equation}
Here $m$ is the adsorbate mass, $\theta$ the coverage, $A$ the adlayer unit cell area, $\rho$ the substrate density and 
$c_{T,L}$ the transversal and longitudinal substrate sound velocities. $D$ is the elastic continuum Green's function, 
which has to be evaluated for the only reciprocal lattice vectors of the overlayer structure $\textbf{q}_i$ included in the 
first Brillouin zone of the substrate \cite{Lewis-JCP-1998}: 
\begin{subequations}
\begin{align}
D(c_{T,L},\rho;\textbf{q}_i,\omega) 	& = i \left[ \frac{q_{i,y}^2}{q_i^2\beta p_T} + \frac{q_{i,x}^2\omega^2}{2q_i^2c_T^2} 
                                                              \frac{p_T}{p_L} \frac{1}{Q(q_i^2,\omega)}  \right], \\
p_{L,T} 						& = \left[\frac{ \omega^2}{c_{L,T}^2} - q_i^2 \right] ^{1/2}, \\
Q(q_i^2,\omega)				& = 2\beta q_i^2 p_T + \frac{p_T^2-q_i^2}{2p_L}\left[ (\alpha - \beta)q_i^2 + (\alpha + \beta )p_L^2\right], \\
\alpha						& = \rho (c_L^2 - c_T^2),\\
\beta							& = \rho c_T^2,
\end{align}
\end{subequations}
where $q_i^2 = \left|\textbf{q}_i\right|^2 = q_{i,x}^2 + q_{i,y}^2$.
Note that an alternative derivation of the ECM for ordered-adlayers has been 
later proposed by Persson and coworkers and generalized to finite-adsorbate size \cite{Persson-PRB-1999}. 

The elastic continuum approximation is meant to be valid for adsorbate modes whose frequency
falls within the long-wavelength tail of the phonon spectrum and for flat surfaces of close-packed materials  
\cite{Lewis-JCP-1998, Pykhtin-PRL-1998}. Both conditions are satisfied for the FT-mode
of Na adsorbed on a low-index surface of copper (Cu(100)), justifying the application of the ECM 
to this system, as also carried out in Refs. \cite{Graham-SurfSciRep-2003, Senet-ChemPhysLett-1999}.

\section{Molecular Dynamics Calculations \label{sec:dynamics}}

\subsection{Preparation of initial conditions\label{sub:initial_conditions}} 

The initial conditions of the molecular dynamics (MD) simulations
presented in the main paper have been prepared according to the following procedure. As
described by Estreicher and coworkers \cite{West-PRL-2006, West-PRB-2007, 
Gibbons-PRL-2009, Estreicher-ModellingSimulMaterSciEng-2009},
the initial energy $E^0(\textbf{q}_n,b)$ assigned to the phonon mode
with wave vector $\textbf{q}_n$ and the band index $b$
is set from an inverse transform 
sampling that allows one to generate Boltzmann-distributed energies at the desired temperature $T$:
\begin{equation}
E^0(\textbf{q}_n,b) = -k_BT\ln\left(1-\zeta(\textbf{q}_n,b)\right).  \label{eq:InitialEnergy}
\end{equation} 
Here  $k_B$ is the Boltzmann 
constant and $\zeta(\textbf{q}_n,b)$ is a random number in the range $\left[0,1\right]$. 
The initial energy of all the adsorbate and substrate modes
is generated according to Equation \ref{eq:InitialEnergy}, 
except for the initial conditions of the Na adlayer which are described at the end of this section.
The initial amplitude of each phonon mode is then calculated as:
\begin{equation}
A^0(\textbf{q}_n,b) = \sqrt{\frac{2E^0(\textbf{q}_n,b)}{\left[\omega(\textbf{q}_n,b)\right]^2}},  \label{eq:Amplitude}
\end{equation} 
where $\omega(\textbf{q}_n,b)$ is the mode frequency.
A second set of random numbers $\{\phi^0(\textbf{q}_n,b)\}_{\textbf{q}_n,b}\in\left[0,2\pi\right)$ is used 
to set the initial phase of each phonon mode.
We have derived the following expressions for the phonon expansion coefficients 
$\{C(\textbf{q}_n,b;t)\}_{\textbf{q}_n,b}$ and their time derivatives 
$\{\dot{C}(\textbf{q}_n,b;t)\}_{\textbf{q}_n,b}$:
\begin{subequations}
\begin{align}
C(\textbf{q}_n,b;t)		&= \frac{1}{2}\left\{ A^0(\textbf{q}_n,b)e^{-i\left[\phi^0(\textbf{q}_n,b)-\omega(\textbf{q}_n,b)t\right]} + 
A^0(-\textbf{q}_n,b)e^{i\left[\phi^0(-\textbf{q}_n,b)-\omega(-\textbf{q}_n,b)t\right]} \right\}, \label{eq:coeff}\\
\dot{C}(\textbf{q}_n,b;t)	&= \frac{i\omega(\textbf{q}_n,b)}{2}\left\{ -A^0(\textbf{q}_n,b)e^{-i\left[\phi^0(\textbf{q}_n,b)-\omega(\textbf{q}_n,b)t\right]} + 
A^0(-\textbf{q}_n,b)e^{i\left[\phi^0(-\textbf{q}_n,b)-\omega(-\textbf{q}_n,b)t\right]} \right\}, \label{eq:dcoeff}
\end{align}
\end{subequations}
which can be easily translated into real-space atomic displacements and velocities. Note that by setting $t=0$,
Equations \ref{eq:coeff} and \ref{eq:dcoeff} provide the displacements 
and the velocities that are employed as initial conditions in the MD runs.
Furthermore, Equations \ref{eq:coeff} and \ref{eq:dcoeff} represent the analytical 
solution for the time evolution of the system based on its phonon (i.e. harmonic) expansion. 

Two types of MD simulations are described in the main paper: In the first type of calculations, in each system we initially excite a phonon mode at $\bar{\Gamma}$ that is localized in the adsorbate overlayer as probed by experiments like HAS. In order to obtain the corresponding phonon mode eigenvectors, we diagonalize the dynamical matrix of the fully-coupled adsorbate-substrate system in the limit of setting the masses of the substrate atoms to infinity. Note that this localized basis of phonon eigenvectors for the adsorbate overlayer subsystem is only used to setup the initial conditions and to calculate the energy still localized on the initially excited adsorbate mode as a function of time (i.e. Figs~\ref{fig:num_layers} to \ref{fig:initial_energy}). The time propagation is always carried out in the fully coupled system with mobile substrate atoms.

The second type of simulations are based on the phonon mode eigenvectors that result from diagonalization of the conventional dynamical matrix of the full system with regular atomic masses. We initially excite the phonon mode  at $\bar{\Gamma}$ described by the eigenvector from this basis that has the largest overlap with the eigenvector initially excited in the first type of simulations for the corresponding system size. This allows us to quantify the anharmonic contribution of the adsorbate-mode lifetime as a function of the temperature as used in Fig. 2 in the the main paper.

\subsection{Phonon Projection Analysis\label{sub:phonon_projections}} 

For each trajectory we calculate the energy corresponding to the Na FT-mode 
at constant time intervals (every 50 fs). This is carried out through the phonon projection 
scheme described in Refs. \cite{McGaughey-PRB-2004, Meyer-PhDThesis,
Bukas-PRL-2016, Bukas-JCP-2017}. 
Briefly, the following transformation relates the real space 
atomic displacements $\{\mathbf{U}_I(t)\}_I$ to the expansion coefficients 
in the phonon basis $\{C(\mathbf{q}_n,b;t)\}_{\mathbf{q}_n,b}$: 
\begin{equation}
C(\mathbf{q}_n,b;t) = \frac{1}{\sqrt{N}}\sum_{I=1}^{N_I} \sqrt{M_I} \left[\mathbf{u}_{\tilde{I}(I)}(\textbf{q}_n,b)\right]^* \cdot \mathbf{U}_I(t) e^{-i\mathbf{q}\cdot\mathbf{R}_I}.
\label{eq:phonon_coefficients}
\end{equation}
Here $C(\textbf{q}_n,b;t)$ is the (complex valued) expansion coefficient corresponding 
to the phonon mode with wave vector $\textbf{q}_n$ and band index $b$. 
$M_I$ and $\mathbf{R}_I$ are the mass and the equilibrium position of the atom $I$, respectively. 
The sum in Eq. \ref{eq:phonon_coefficients} runs over all the $N_I$ atoms in the systems.
$N$ is the number of replicas of the primitive cell that are included into the simulation cell.
The map $\tilde{I}(I)$ associates to the atom $I$
its corresponding image inside the primitive cell and 
$\mathbf{u}(\textbf{q}_n,b)$ is the $b$-th eigenvector of the dynamical matrix of the system, 
evaluated at $\textbf{q}_n$.
These eigenvectors and the corresponding phonon frequencies $\{\omega(\textbf{q}_n, b)\}_{\mathbf{q}_n,b}$ 
have been calculated using the finite difference method \cite{Parlinski-PRL-1997} as implemented in  
PHONOPY \cite{Togo-ScriptaMaterialia-2015}. 
Note that by substituting the atomic velocities $\{\dot{\mathbf{U}}_I(t)\}_I$ to the atomic 
displacements $\{\mathbf{U}_I(t)\}_I$ in Eq. \ref{eq:phonon_coefficients} one 
obtains the analogous expression for the time derivatives of the phonon coefficients, i.e. 
$\{\dot{C}(\mathbf{q}_n,b;t)\}_{\mathbf{q}_n,b}$. 

The coefficients $\{C(\mathbf{q}_n,b;t)\}_{\mathbf{q}_n,b}$ carry information about 
both the amplitude and the phase of each phonon mode. From them (from their
time derivatives) one can straightforwardly calculate the phonon 
mode-specific potential (kinetic) energy as a function of time.
The time-resolved 
and phonon mode specific energy $E(\mathbf{q}_n,b;t)$ is calculated as:
\begin{equation}
E(\mathbf{q}_n,b;t) =  V(\mathbf{q}_n,b;t) + K(\mathbf{q}_n,b;t) =  
\frac{1}{2}\omega^2(\mathbf{q}_n,b)\left|C(\mathbf{q}_n,b;t)\right|^2 + 
\frac{1}{2}\left|\dot{C}(\mathbf{q}_n,b;t)\right|^2,
\end{equation}
where $V(\mathbf{q}_n,b;t)$ and $K(\mathbf{q}_n,b;t)$ are the phonon mode
specific potential and kinetic energy, respectively.   
The mode specific coefficient $C(\mathbf{q}_n,b;t)$ and its time 
derivative $\dot{C}(\mathbf{q}_n,b;t)$ can be calculated from the real space 
atomic displacements and velocities, respectively, using Eq. \ref{eq:phonon_coefficients} and its analogous expression 
for time derivatives.
The code employed to perform this phonon projection analysis  
is available upon request to the authors.

Vibrational lifetimes have been extracted from MD calculations 
by fitting an exponential decay function to the average energy in the Na FT-mode
as a function of time:
\begin{equation}
E_{\mathrm{FT}}(t) = (E_i-k_BT)\cdot e^{-t/\tau} + k_BT,
\label{eq:exp_decay}
\end{equation}
according to which $E_{\mathrm{FT}}(t=0)=E_i$
and $E_{\mathrm{FT}}(t\rightarrow\infty)=k_BT$, as expected for sufficiently large
system sizes.  

Even though the thermal energies considered here are small compared to the Na diffusion 
barrier ($k_BT=6.5$ meV at 75 K), thermal fluctuations can give rise to
trajectories where Na atoms hop to neighboring hollow sites. 
The trajectories characterized by these minority events cannot be employed 
for the purpose of studying the de-excitation of the Na FT-mode and are therefore 
discarded for our analysis.

\subsection{Finite Size and Convergence Tests\label{sub:convergence}} 

Due to the finite size of the system, phonon waves emitted from the adsorbates 
will be reflected at the boundaries of the cell and provoke a recursive energy increase 
in the monitored adsorbate mode. The time delay with which this recursion is 
observed is proportional to the thickness of the slab. We have found that by 
discarding the times $t > t_c$ with $t_c = 2L/c_T$, where $L$ is the slab thickness and $c_T$ the
transversal speed of sound in Cu along the [100] direction, we obtain stable 
estimates of the lifetime (see following section). $t_c$ represents the time required by
a transversal phonon wave (as likely emitted by a phonon mode with parallel polarization
vector such as the Na FT-mode) to travel twice across the slab. 
For a slab thickness corresponding to 24 atomic layers $t_c \approx 2.9$ ps. 

\begin{figure}[H]
\begin{centering}
	\includegraphics[width=0.7\columnwidth]{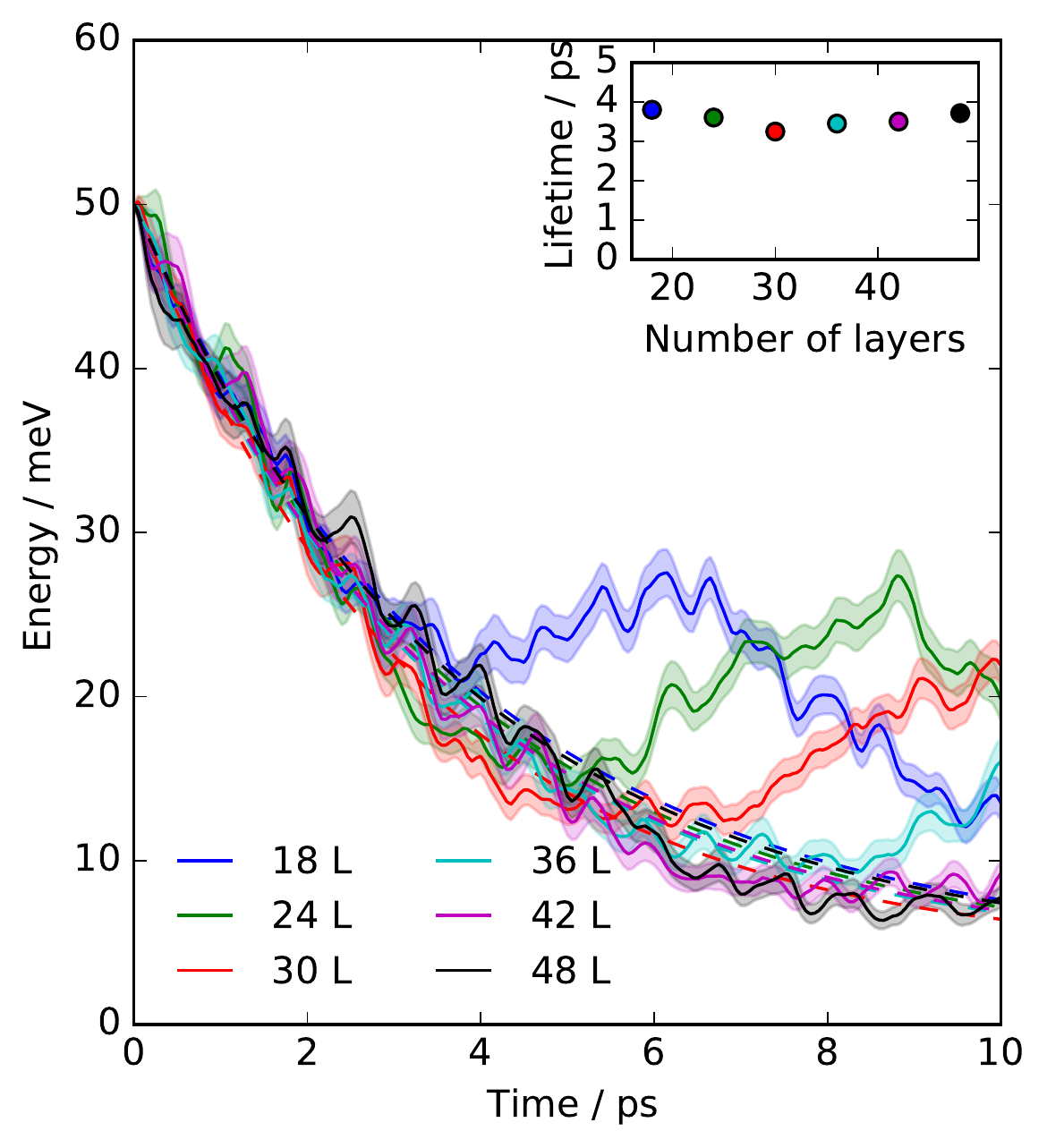}\\
\end{centering}
\caption{Average energy in the Na FT-mode as a function of time for various number of layers (solid lines).
The shaded areas around the curves include energies within one $\sigma$ (standard error of the mean) 
from the average. The temperature $T=50$ K has been simulated here. Dashed lines corresponds to 
exponential decay functions (see Equation \ref{eq:exp_decay}) fitted to the data. The inset shows the lifetimes
extracted from these fits for the various number of layers.}
\label{fig:num_layers}
\end{figure}

\begin{figure}[H]
\begin{centering}
	\includegraphics[width=0.7\columnwidth]{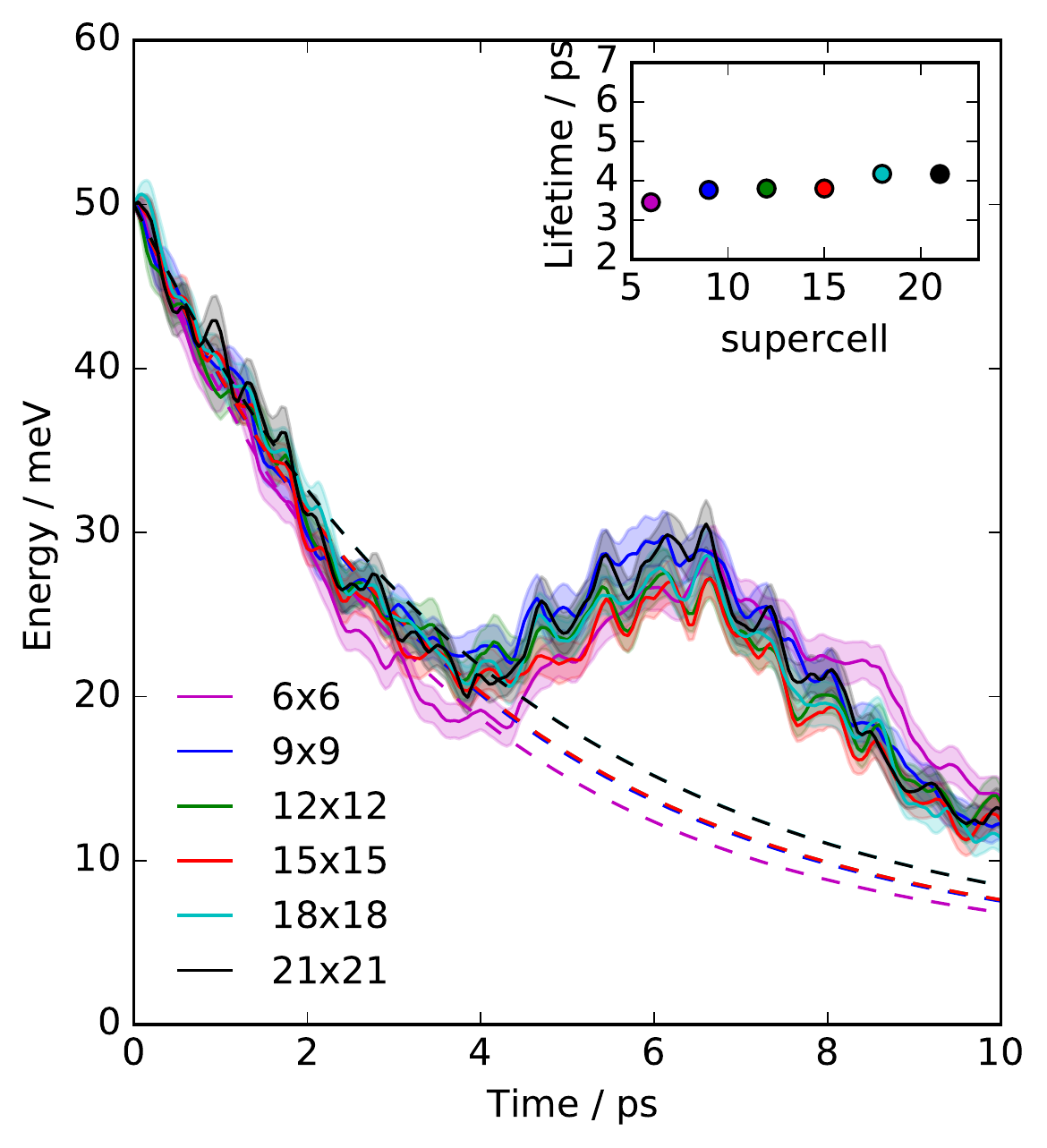}\\
\end{centering}
\caption{Same as Figure \ref{fig:num_layers}, but varying the lateral extension of the supercell employed
 (expressed in terms of the substrate supercell). The temperature $T=50$ K has been simulated here, 
 using a 18-layer thick slab.}
\label{fig:supercell}
\end{figure}

\begin{figure}[H]
\begin{centering}
	\includegraphics[width=0.7\columnwidth]{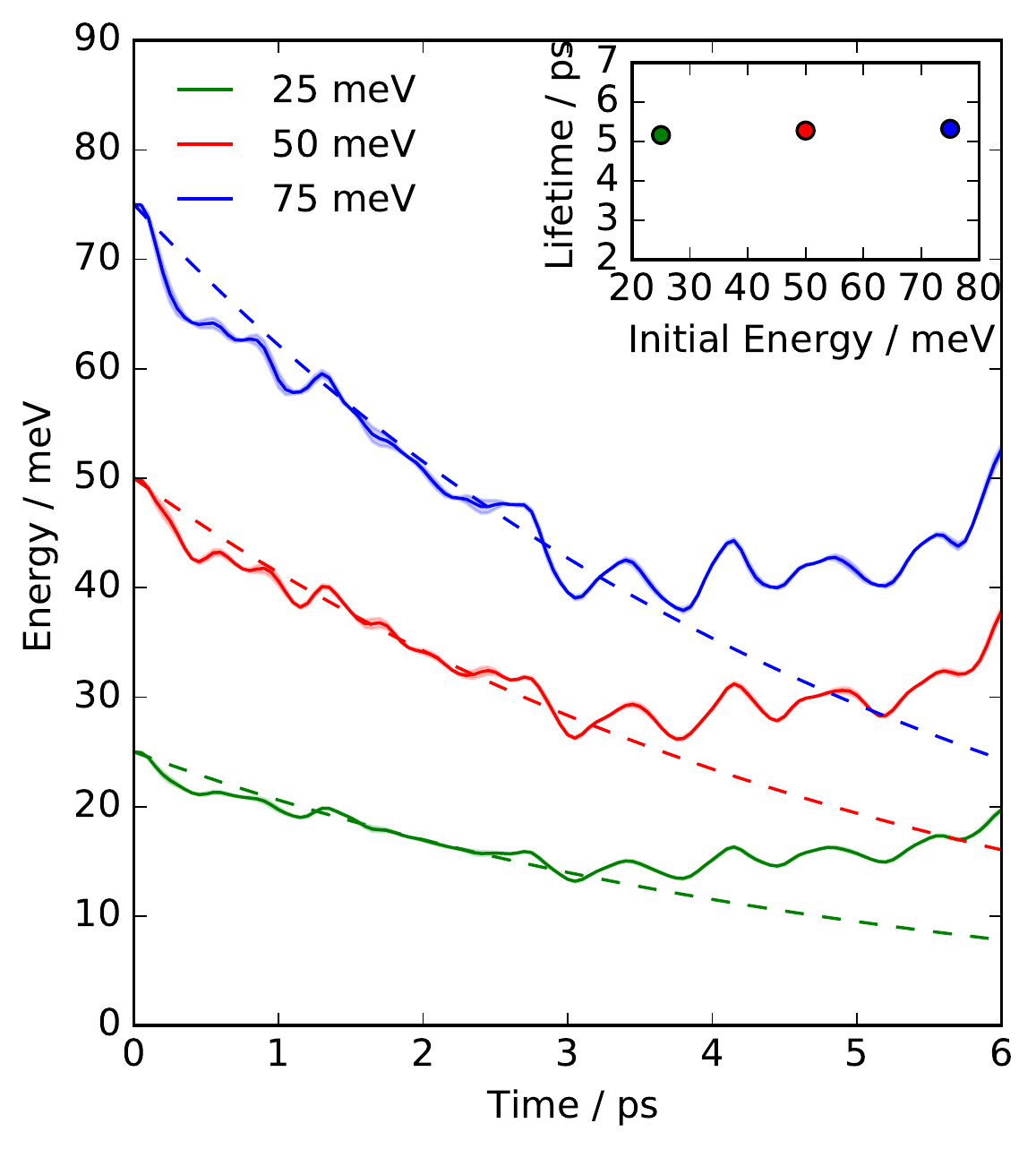}\\
\end{centering}
\caption{Same as Figure \ref{fig:num_layers}, but varying the initial excitation energy in the Na FT-mode. 
The temperature $T=0$ K has been simulated here, using a 24-layer thick slab.}
\label{fig:initial_energy}
\end{figure}

We have carefully checked the convergence of the computational setup  
employed with respect to the chosen computational parameters. 
All calculations described in this section have been performed 
for one representative overlayer structure, namely the p$(3\times3)$ structure. 
All calculations described in this section have been performed in a supercell corresponding to a 
12x12 multiple of the substrate primitive cell with an initial excitation energy of the Na FT-mode
corresponding to 50 meV, if not otherwise specified.
Increasing the thickness of the slab from 18 to 48 atomic layers does not 
affect the estimate of the lifetime, as shown in Figure \ref{fig:num_layers}. 
Note that even though recursions due to the finite size of the cell are clearly visible, 
the lifetime determination is stable, as we discard the times $t > t_c$ which
might be affected by artificial phonon reflection from the 
boundaries of the cell (see also previous paragraph). 
Similarly, increasing the supercell size or varying the 
initial excitation energy in the Na FT-mode does not affect the lifetime determination,
as illustrated in Figure \ref{fig:supercell} and \ref{fig:initial_energy}, respectively. 

\subsection{Electronic Friction Model \label{sub:electronic_friction}}

Electronic friction dynamics calculations are carried out here 
in order to account for electron-hole pair excitations 
\cite{Tully-JCP-1980, Tully-JVacSciTechnol-1993, HeadGordon-JCP-1995, Juaristi-PRL-2008}.
We apply here a constant friction coefficient, as suitable for describing the 
friction force acting on sodium atoms undergoing small displacements from their 
equilibrium positions. 

\begin{figure}[H]
\begin{centering}
	\includegraphics[width=0.7\columnwidth]{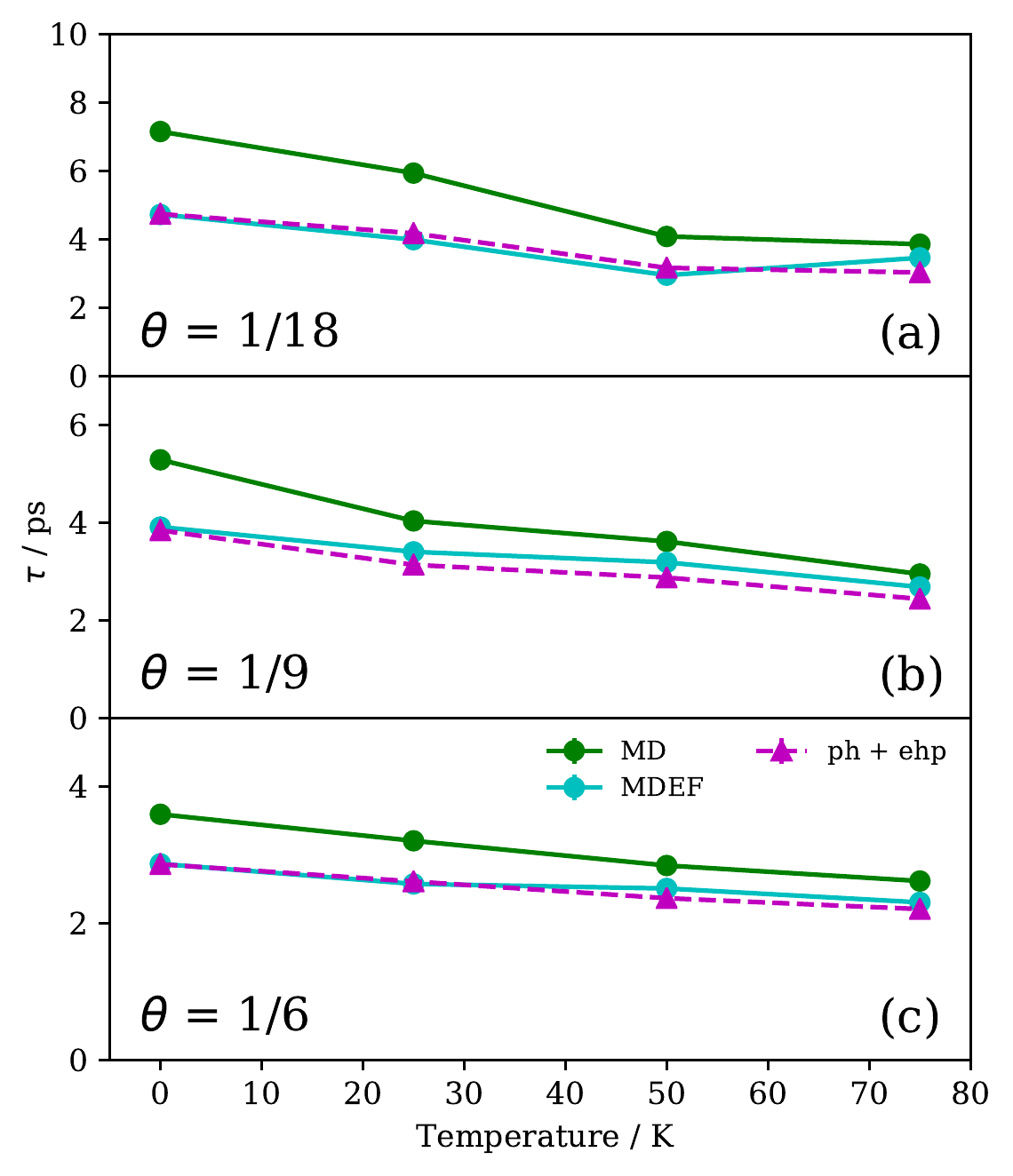}\\
\end{centering}
\caption{Same as Figure 3 in the main paper, but the constant friction coefficient
applied here has been calculated using a different LDFA implementation \cite{Rittmeyer-PRL-2016}, 
i.e. the LDFA-IAA approach. 
The Na FT-mode lifetimes are shown as a function of the temperature for the 
various overlayer structures (panels (a) to (c)). The 
lifetimes obtained through MD calculations (green symbols, solid line to guide the eyes)
are compared to the lifetimes obtained from MDEF calculations (cyan symbols,
 solid line to guide the eyes) and to 
the lifetimes estimated from the sum of the phononic and electronic contributions 
(ph + ehp, violet symbols, dashed line to guide the eyes). 
Error bars reflect the quality of the fits employed to estimate the lifetimes.}
\label{fig:electronic_friction}
\end{figure}

Considering that we do not aim at producing quantitatively accurate estimates 
of electronic non-adiabatic effects on the lifetime of the FT-mode of Na adsorbed 
on Cu(100), but we rather investigate the additivity of electronic and phononic 
dissipation mechanisms, we make use of the average friction coefficient $\eta_{\mathrm{ehp}}$
as determined in Ref. \cite{Rittmeyer-PRL-2016} for Na diffusing on Cu(111).
Two values of $\eta_{\mathrm{ehp}}$ are reported in Ref. \cite{Rittmeyer-PRL-2016}, as calculated 
on the basis of two implementations of the local density friction approximation (LDFA).
These implementations differ in the choice of the adsorbate embedding 
density, quantity to which $\eta_{\mathrm{ehp}}$ is proportional:
the independent-atom-approximation (IAA) \cite{Juaristi-PRL-2008} makes use of the electron density of the clean surface,
while the atom-in-molecule (AIM) approach \cite{Rittmeyer-PRL-2015} accounts for the density modifications due to
the presence of the adsorbate (e.g. partial charge transfer). 

Results presented in Figure 3 of the main paper are based on the AIM  
estimate of $\eta_{\mathrm{ehp}}$ ($\eta_{\mathrm{AIM}} = 2.6$ amu/ps \cite{Rittmeyer-PRL-2016}). 
If the molecular dynamics with electronic friction (MDEF) calculations are instead based on 
the IAA-based friction value ($\eta_{\mathrm{IAA}} = 1.64$ amu/ps \cite{Rittmeyer-PRL-2016}),
we calculate lifetimes as reported in Figure \ref{fig:electronic_friction}. 
Note that MDEF lifetimes are closer to the MD values when using the LDFA-IAA model 
for the friction. This is consistent with $\eta_{\mathrm{IAA}}$ being smaller than 
$\eta_{\mathrm{AIM}}$, as a consequence of the larger embedding density predicted by the AIM model
due to the partial positive charge present on Na adsorbates \cite{Rittmeyer-PRL-2016}.   

Figure \ref{fig:electronic_friction} and Figure 3 of the main paper suggest that for the current system  
the additivity of phononic and electronic dissipation channels does not depend on the 
specific value of the damping parameter. 
In fact, our model based on the sum of the contributions
from the two damping mechanisms (Equation 3 in the main paper) quantitatively 
reproduce the results of MDEF calculations regardless of whether 
IAA- and AIM-based friction coefficients are employed. 

\begin{figure}[H]
\begin{centering}
	\includegraphics[width=0.7\columnwidth]{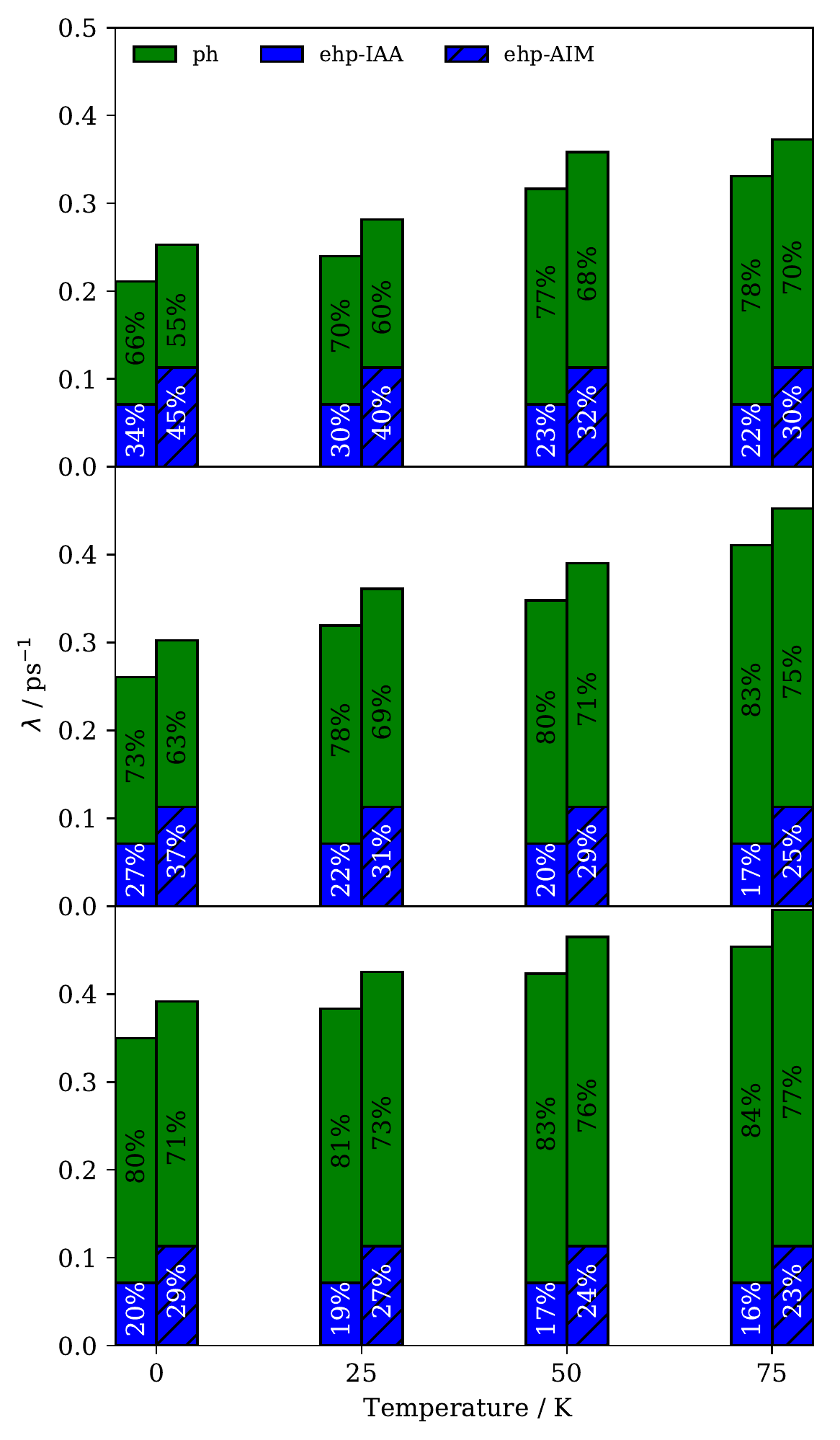}\\
\end{centering}
\caption{Na FT-mode dissipation rates as a function of the temperature for the various overlayer
structures. Each rate is decomposed into the phononic (green) and the electronic (blue) contribution 
(numbers indicate percentage to the overall rate).
The results of IAA and AIM-friction-based models are shown as plain and shaded bars, respectively.}
\label{fig:electronic_friction_comparison}
\end{figure}

After having verified the additivity of the phononic and electronic 
dissipation mechanisms, we finally consider the relative importance of the two 
contributions in the damping of the Na FT-mode. Figure \ref{fig:electronic_friction_comparison} 
shows the relative contributions of the phononic and electronic dissipation channels to the 
overall rate $\lambda\equiv 1 / \tau$. For both the IAA and the AIM friction models, the absolute electronic contribution
is independent on temperature and coverage structure. However, the relative contribution to the overall rate
decreases with increasing temperature, due to the increased phononic rate contribution. 
Nevertheless, at the low temperatures considered in the current work, both the AIM and IAA friction models 
suggest a surprisingly relevant electronic contributiion, in agreement with the recent findings of Rittmeyer et al. 
\cite{Rittmeyer-PRL-2016}. 
 
\bibliographystyle{apsrev4-1} 
\bibliography{Manuscript}